\documentclass[preprint,aps,pre,letterpaper,superscriptaddress,longbibliography]{revtex4-2}
\usepackage{graphicx}
\usepackage{natbib}
\usepackage{amssymb}
\usepackage{amsmath}
\usepackage[usenames]{color}
\usepackage[normalem]{ulem}
\renewcommand{\sout}[1]{} 
\usepackage{subcaption}
\usepackage{epstopdf}

\newcommand{\mbf}[1]{\boldsymbol{#1}}

\newcommand{\balpha}{\boldsymbol{\alpha}}

\newlength{\figwidth}
\newlength{\SCwidth}
\def\XXint#1#2#3{{\setbox0=\hbox{$#1{#2#3}{\int}$}
     \vcenter{\hbox{$#2#3$}}\kern-.5\wd0}}

\newcommand{\AS}[1]{\textcolor{black}{#1}}

\newcommand{\MDGreviseII}[1]{\textcolor{black}{#1}}
\newcommand{\MDGrevise}[1]{\textcolor{black}{#1}}
\newcommand{\Wi}{\mathrm{Wi}}
\newcommand{\Rey}{\mathrm{Re}}

\begin{document}

\title{Tollmien-Schlichting route to elastoinertial turbulence in channel flow}
\author{Ashwin Shekar}
\affiliation{Department of Chemical and Biological Engineering, University of Wisconsin-Madison, Madison WI 53706, USA}

\author{Ryan M. McMullen}
\affiliation{Graduate Aerospace Laboratories, California Institute of Technology, Pasadena CA 91125, USA}

\author{Beverley J. McKeon}
\affiliation{Graduate Aerospace Laboratories, California Institute of Technology, Pasadena CA 91125, USA}

\author{Michael D. Graham}
\email{mdgraham@wisc.edu}
\affiliation{Department of Chemical and Biological Engineering, University of Wisconsin-Madison, Madison WI 53706, USA}

\date{\today}

\begin{abstract}
Direct simulations of two-dimensional channel flow of a viscoelastic fluid have revealed the existence of a family of Tollmien-Schlichting (TS) attractors that is nonlinearly self-sustained by viscoelasticity [Shekar et al., \emph{J.~Fluid Mech.} \textbf{893}, A3 (2020)]. 
Here, we describe the evolution of this branch 
in parameter space and its
connections to the Newtonian TS attractor and to elastoinertial turbulence (EIT). At Reynolds number $\Rey=3000$, there is a solution branch with TS-wave structure but which is not connected to the Newtonian solution branch.  At fixed Weissenberg number, $\Wi$ and increasing Reynolds number from 3000-10000, this attractor goes from displaying a
striation of weak polymer stretch localized at the critical
layer to an extended sheet of very large polymer stretch. We show that this transition is directly tied to the strength of the TS critical layer fluctuations and can be attributed to a coil-stretch transition when the local Weissenberg number at the hyperbolic stagnation point of the Kelvin cat's eye structure of the TS wave exceeds $\frac{1}{2}$. At $\Rey=10000$, unlike $3000$, the Newtonian TS attractor evolves continuously into the EIT state as $\Wi$ is increased from zero to about $13$. We describe how the structure of the flow and stress fields changes, highlighting in particular a ``sheet-shedding" process by which the individual sheets associated with the critical layer structure break up  to form the layered multisheet structure characteristic of EIT.

\end{abstract}

\maketitle   
\newpage

\section{Introduction}

%
%
%
%
%
%

The addition of minute quantities of long chain polymer molecules has tremendous effects on wall-bounded turbulence, the most dramatic and well known effect being the reduction of friction factor by levels unmatched by additive-free turbulence control-schemes. The rheological drag reduction phenomenon was first discovered by Toms in the 1940s \MDGrevise{\citep{Toms_P1INTCRHEOL1948,tomsremin}}, and is sometimes referred to as the ``Toms effect". This technology has found use in various applications that seek energy efficiency -- transportation of oil, fracking and heat $\&$ cooling systems, to name a few \MDGrevise{\citep{Fink:2012,Burger:1982,King:2012vq}.} 

\AS{Accompanying the drastic reduction in friction factor is a structural change to the flow -- a well known effect is the suppression by viscoelasticity of the near-wall streamwise vortices that dominate Newtonian turbulence \citep{Kim:2007dq, White:2008hs}.} \AS{A number of studies have captured this phenomena by studying the effect of viscoelasticity on three-dimensional (3D) nonlinear traveling wave solutions of the Navier-Stokes equations termed exact coherent states (ECS) \citep{Stone:2002dj, Stone:2003gq, Stone:2004jk, Li:2005vl, Li:2006gk, Li:2007ii}.} These ECS contain the basic self-sustaining ingredients of transitional Newtonian turbulence i.e quasistreamwise vortices and streaks. A comprehensive review of Newtonian ECS can be found in \cite{Graham:2020ba}. 
\MDGrevise{In particular,}
 Li and coworkers \cite{Li:2006gk,Li:2007ii}  focused on one such family of ECS in channel flow and observed a weakening of structures on increasing Weissenberg number ($\Wi$), the ratio between the polymer relaxation time scale and the flow time scale. \MDGrevise{At sufficiently high $\Wi$, these structures are so weakened that the ECS are no longer self-sustained and lose existence.} \AS{These results indicate that the near-wall self-sustaining process of Newtonian turbulence is disrupted by viscoelasticity.} \MDGrevise{Recognizing that in general, viscoelasticity is not experimentally observed to drive relaminarization, these authors suggested the possibility of new viscoelastic mechanisms for turbulence coming into existence and being unmasked as the Newtonian structure are suppressed \citep{Li:2006gk}.}

\begin{figure}
	\begin{center}
		\begin{subfigure}{0.48\textwidth}
	    \includegraphics[width=0.9\textwidth]{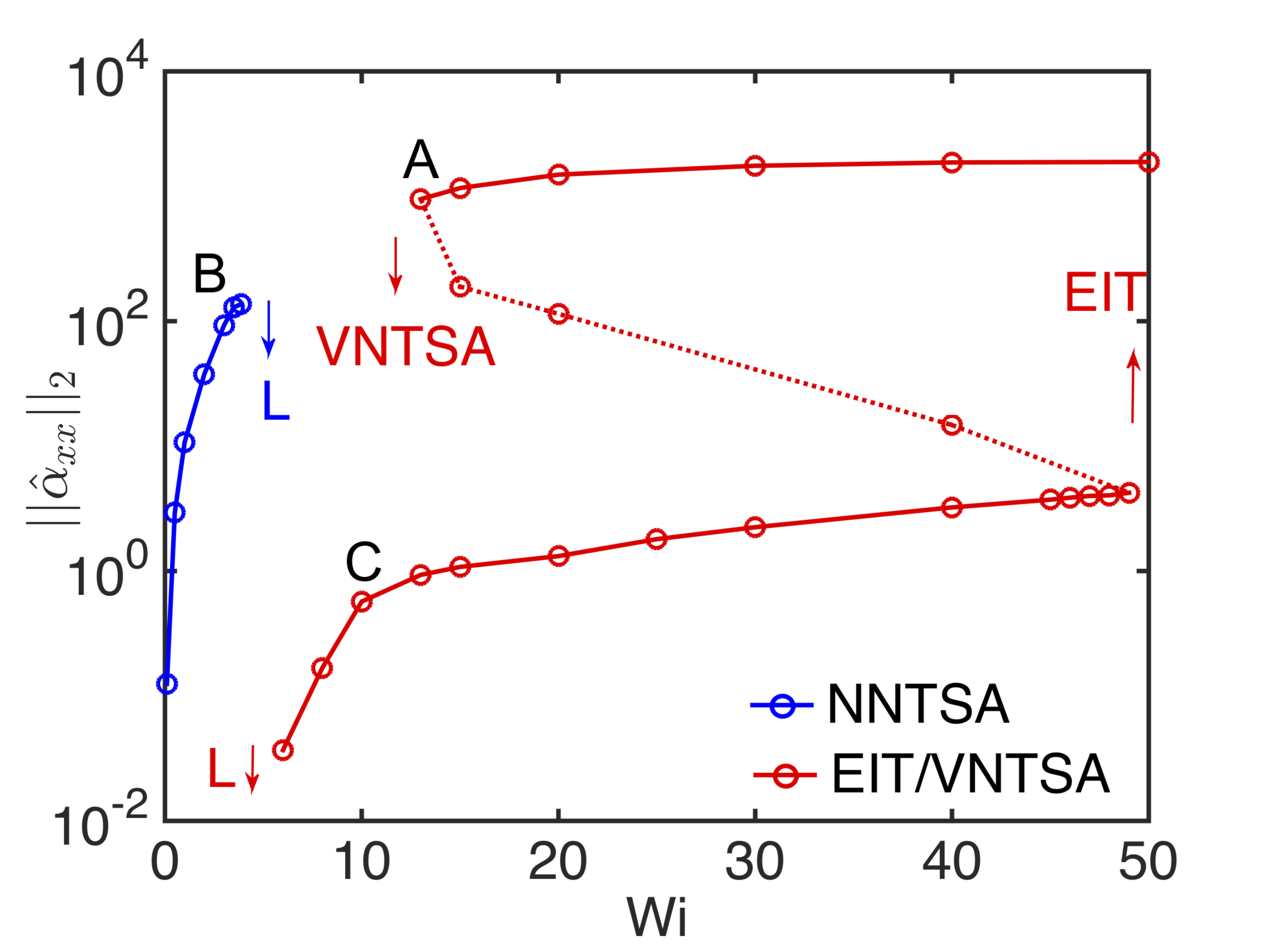} 
	    \caption[]{}
      	\label{fig:BFD_3000}
        \end{subfigure}
		\begin{subfigure}{0.48\textwidth}
	    \includegraphics[scale=0.36]{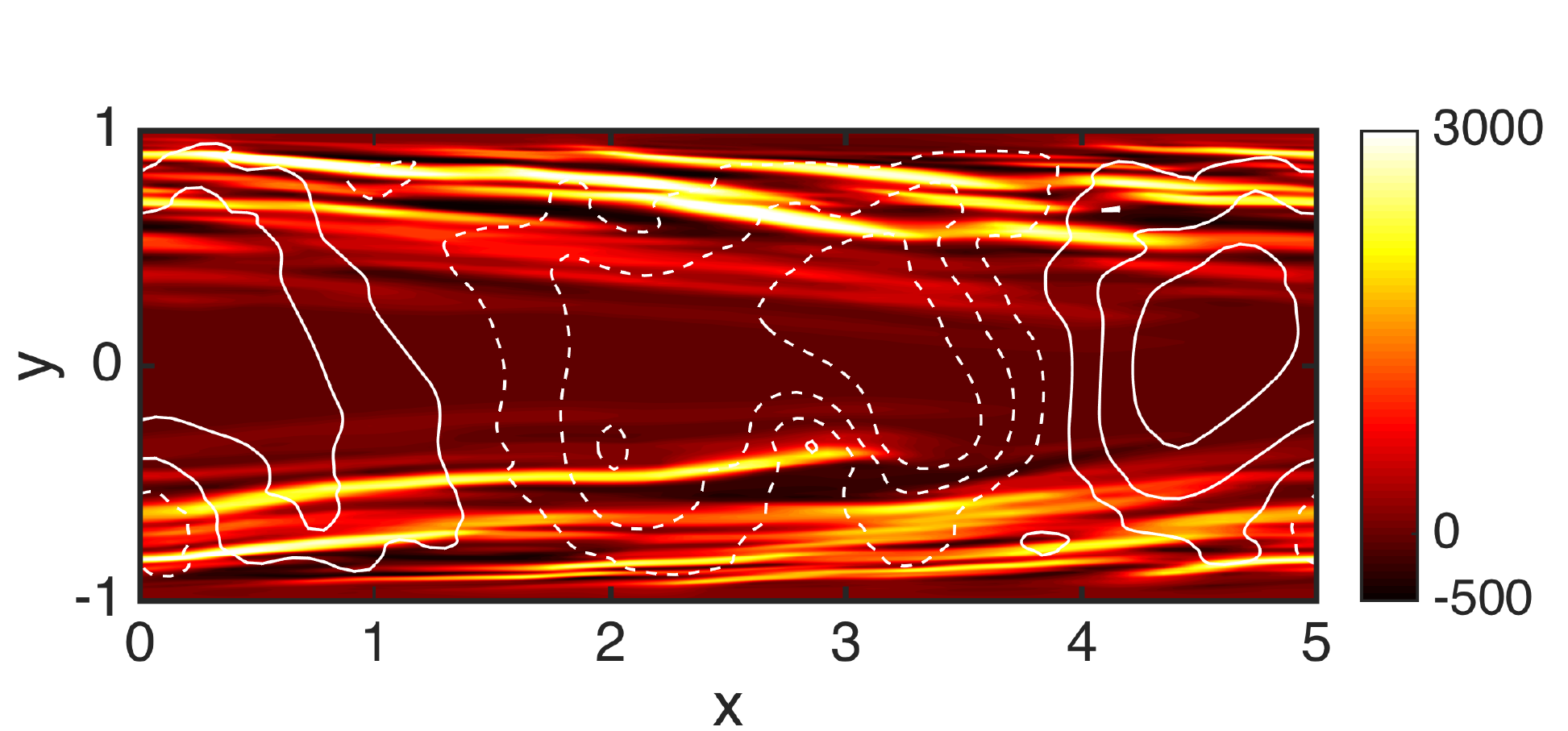} 
	    \caption[]{}
      	\label{fig:EIT_Re3000_Wi13}
        \end{subfigure}
		\begin{subfigure}{0.48\textwidth}
	    \includegraphics[scale=0.29]{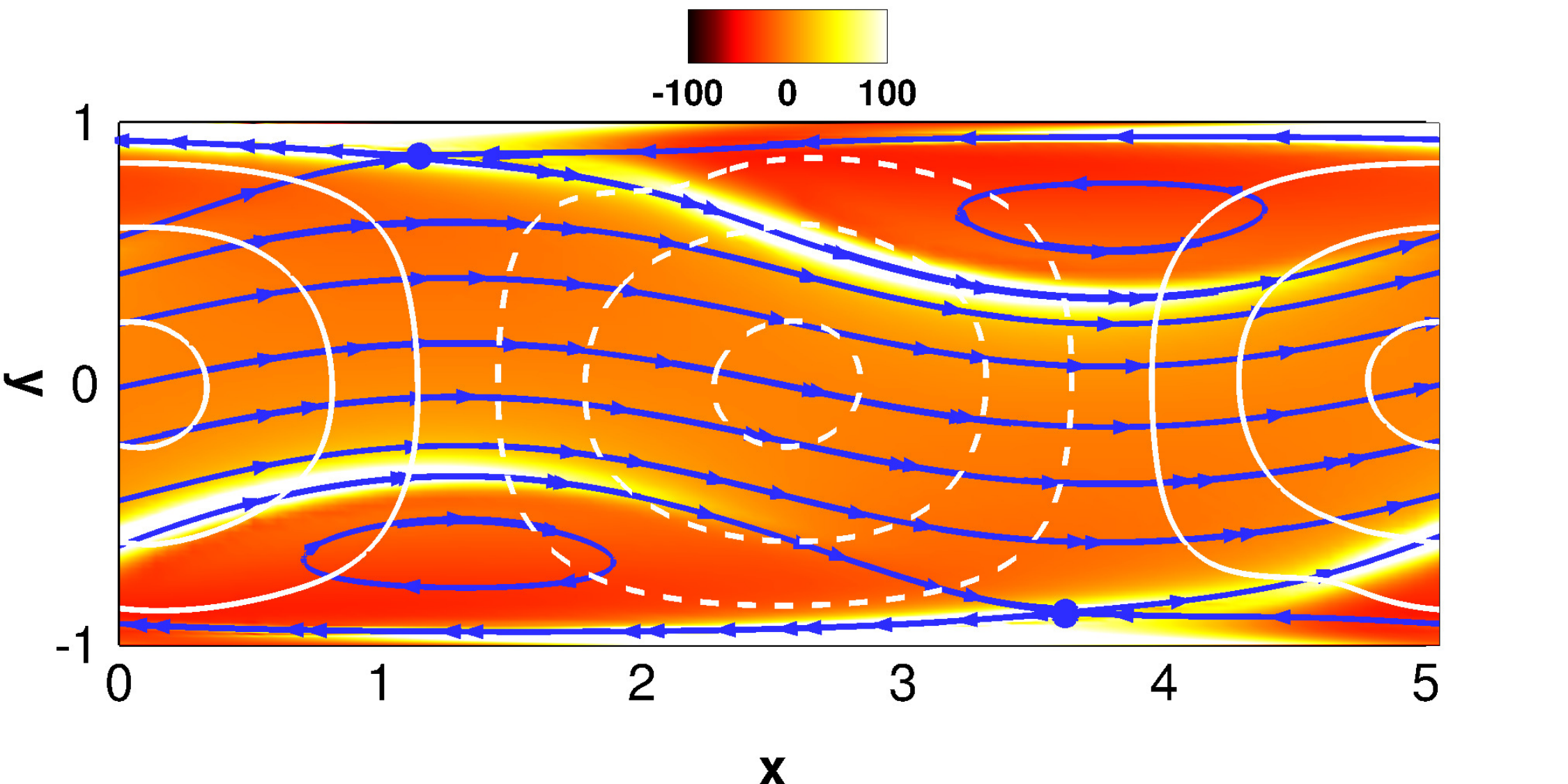} 
	    \caption[]{}
      	\label{fig:NNTSA_Re3000_Wi3}
        \end{subfigure}
		\begin{subfigure}{0.48\textwidth}
		\includegraphics[scale=0.36]{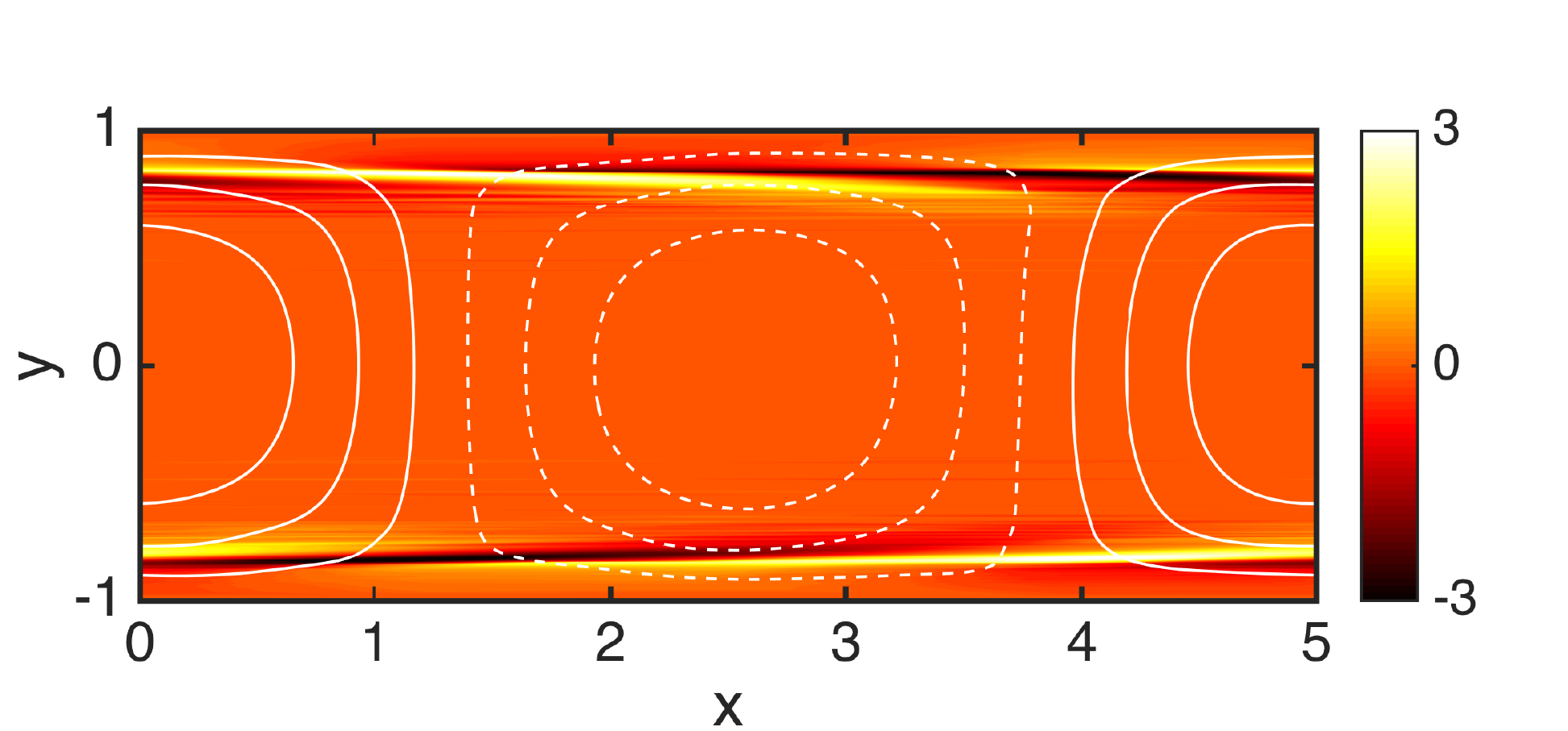} 
		\caption[]{}
		\label{VNTSA_Re3000_Wi10}
		\end{subfigure} 
\end{center}
\caption[]{\AS{(a) Time-averaged $L_2$-norm of $\hat{\alpha}_{xx}$ vs $\Wi$ of the three solution branches identified in \cite{Shekar:2020gt} at $\Rey = 3000, L_x = 5$.} Snapshots corresponding to (b)  2D EIT at $\Wi = 13$ (point A on the bifurcation diagram); (c)  NNTSA at $\Wi = 3$ (point B); and (d)  VNTSA at $\Wi = 10$ (point C). Shown are white contour lines of wall normal velocity, $\hat{v}$ superimposed on color contours of xx-component of polymer conformation tensor ($\hat{\alpha}_{xx}$). Here $\hat{}$ denotes deviations from laminar base state. For point B, we also show the streamlines (blue) in a reference frame moving at the wave speed and the hyperbolic stagnation points (blue dots).}
\label{fig:JFM_2020}	     
\end{figure} 

\MDGrevise{Indeed}, instead of complete relaminarization (except in narrow parameter ranges at transitional $\Rey$ as detailed later), recent studies have unearthed a polymer-driven chaotic flow state dubbed elastoinertial turbulence (EIT) dominating the flow at high levels of viscoelasticity \citep{Samanta:2013el}. EIT was shown to display tilted multilayered sheets of polymer stretch with weak spanwise-oriented flow structures -- a sharp contrast to the 3D quasistreamwise vortex structures that make up inertia-driven Newtonian turbulence. Gaining an understanding of the self-sustenance mechanisms that underlie EIT has been the focus of recent efforts. \AS{Here, we build on previous observations that tie EIT to mechanisms originating from Tollmien-Schlichting (TS) waves and uncover the underlying process that gives rise to the structures seen in EIT.}

Choueiri et al.~\cite{Choueiri:2018it} experimentally examined EIT by focusing on parameter regimes close to the inception of Newtonian turbulence in pipes. At low transitional $\Rey = 3150$ and varying polymer concentration in a quasi-static way, these authors observed a complete relaminarization of Newtonian turbulence in a narrow window of concentration, followed by a reentrant transition to EIT. These observations bring to light two distinct self-sustenance mechanisms in dilute polymer solutions, one that is suppressed by viscoelasticity (Newtonian turbulence) and one that is promoted (elastoinertial turbulence). 

\AS{Using computations of flowing dilute polymeric solutions in channel flow, Shekar et al.~\cite{Shekar:2019hq} shed critical insights into the viscoelastic mechanisms underlying EIT.} At fixed $\Rey = 1500$ and varying $\Wi$, they observed a narrow zone of $\Wi$ where the only attractor was the laminar base state. This separated drag-reduced Newtonian turbulence at lower $\Wi$ and EIT at higher $\Wi$, \MDGrevise{corroborating the experimental } observations of \cite{Choueiri:2018it}. Further, finite amplitude perturbations were required to transition to EIT, \MDGrevise{indicating a}  new viscoelastic bypass transition route to EIT. By analyzing the structures at EIT, they were able uncover clues into the nature of this transition.
\AS{Close to its inception, EIT in this parameter regime  displayed polymer stretch fluctuations localized near the wall.} \AS{To shed light on the origins of these  localized fluctuations, Shekar et al.~\cite{Shekar:2019hq}  used a combination of linear stability analysis and resolvent analysis, which quantifies the linear response of the system to external forcing in frequency space.} \MDGrevise{In particular,} they observed a clear resemblance \MDGrevise{of the EIT structure in this parameter range} to the viscoelastic Tollmien-Schlichting (TS) mode, which at the chosen parameters was the slowest decaying mode from linear stability analysis. Similarly, resolvent analysis predicted strong amplification \MDGrevise{of this structure} in the presence of viscoelasticity. This strong amplification implies that even very weak disturbances may be sufficient to trigger the nonlinear effects necessary to sustain EIT. \MDGrevise{The viscoelastic TS mode displays polymer stretch fluctuations that are sharply localized to critical layers, i.e wall-normal positions near the top and bottom walls where the streamwise velocity equals the real part of the wave speed.}
 \AS{Critical layers can be thought of as the most favorable positions for energy exchange between the mean and fluctuations, because they are the positions where both have the same speed. These results indicate a role for TS-like critical layer mechanisms at EIT.} \MDGrevise{Indeed, the initial EIT computations of Samanta et al.~\cite{Samanta:2013el}, as well as followup computations \citep{terrapon2013dynamics,Terrapon:2014kn} show strong localization of fluctuations in a layer near the wall.} \AS{Sid et al.~\cite{Sid:2018gh} later showed that EIT is fundamentally 2D in nature by performing simulations in 2D channel flows.} Relatedly, Haward  et al.~\cite{Haward:2018he,Haward:2018hs} present experiments and analysis for viscoelastic flow over a wavy wall that illustrate amplification of perturbations in the critical layer.

\AS{Similar structures have been observed in \MDGrevise{a range of}  flow scenarios, indicating a robustness in the underlying mechanisms.} \AS{Near-wall localized, nearly-axisymmetric vortex \MDGrevise{and stress} structures have been reported in pipe flow simulations of EIT by Lopez et al.~\cite{lopez2019dynamics}. } \AS{This near-wall nature of EIT in pipes was recently corroborated in experiments by Choueiri et al.~ \cite{choueiri2021experimental}.} \AS{This is exactly what we would expect, if critical layers are also a strong source of linear amplification in pipe flow, as they are in channel flow \citep{Mckeon:2010ep}.}  \MDGrevise{Indeed, Zhang \cite{Zhang:2021ef} recently performed resolvent analysis for pipe flow in the same parameter regime, demonstrating that the most amplified mode has strong stress fluctuations localized in a critical layer near the wall, just as was found by Shekar et al. \cite{Shekar:2019hq}. }  \AS{Simulations of turbulent Couette flow by Pereira et al.~\cite{pereira2019beyond} indicate a ``pseudo-laminar state" with weak wall-localized fluctuations as a precursor to the emergence of EIT. } \AS{Again, the structures observed were nearly 2D, with near-wall localization.}  \AS{Very recent computations by Zhu et al.~\cite{zhu2021nonasymptotic} observe an intermittent process in drag-reduced turbulence involving spanwise-oriented quasi-2D structures with critical layer characteristics and 3D quasistreamwise structures.} \AS{While the TS instability per se does not exist in \MDGrevise{pipe and plane Couette geometries}, that fact does not preclude nonlinear mechanisms with near-wall structure coming into existence at a finite amplitude to underlie EIT in these contexts.} \AS{We expand on this point below.}

\AS{Given the 2D nature of EIT at transitional $\Rey$, Shekar et al.~\cite{Shekar:2020gt} studied the bifurcation scenario of 2D channel flow EIT \MDGrevise{at $\Rey=3000$} and its ties to nonlinear self-sustaining TS waves.} \MDGrevise{In that study, the domain was chosen to be five channel half-heights long, because at this length, self-sustaining nonlinear Tollmien-Schlichting waves exist in the Newtonian case for $\Rey\gtrsim 2800$ \citep{Jimenez:1990cn}.} \AS{\MDGrevise{Some key results of this study that form the basis for the present work are described below and summarized in} Figure \ref{fig:JFM_2020}}.  \AS{At this parameter set, 2D EIT \MDGrevise{(upper branch of the red curve on the bifurcation diagram of Figure \ref{fig:JFM_2020}a) comes into existence at $\Wi \approx 13$.}} \AS{A snapshot of the \MDGrevise{$xx$-component of the polymer stretch, displaying} tilted, \textit{multilayered sheets} at EIT is shown in Figure \ref{fig:JFM_2020}b.}  \MDGrevise{These authors identify two separate branches of attractors based on the TS mode. The first, (solid blue curve on Figure \ref{fig:JFM_2020}a),  originates with the Newtonian nonlinear traveling wave solution and loses existence above $\Wi \approx 4$. This solution branch is termed the Newtonian nonlinear Tollmien-Schlichting attractor (NNTSA); it develops a {sheet} of high polymer stretch that starts out near the wall and arches away from it as shown in Figure \ref{fig:JFM_2020}c. This structure originates with the Kelvin cat's eye structure of TS waves, specifically its near-wall hyperbolic stagnation points (blue circles in Figure \ref{fig:JFM_2020}c). These observations demonstrate the capability of nonlinear TS critical layer mechanisms in generating sheets of polymer stretch.} 

\MDGrevise{The second solution branch (lower solid-red curve) with TS-like structure comes into existence at finite amplitude when $\Wi \approx 6$. This branch, denoted the viscoelastic nonlinear Tollmien-Schlichting attractor (VNTSA), was first found by starting with an initial condition at EIT, then lowering $\Wi$ below the threshold value $\Wi\approx 13$ noted above. It has extremely weak fluctuations and seems likely to be related to the ``pseudo-laminar" state observed by Pereira et al.~\cite{pereira2019beyond}. This state displays {striations} of polymer stretch  localized at the critical layer (figure \ref{fig:JFM_2020}d) and is structurally very similar to the linear TS mode at the same value of $\Wi$, though it displays weakly chaotic rather than purely time-periodic (traveling wave) fluctuations.}
\AS{The existence of such a state indicates the presence of a nonlinear \MDGrevise{viscoelastic} mechanism for \MDGrevise{self-sustenance of} TS structure, \MDGrevise{an observation that has important implications} in the broader context of other flow scenarios such as pipe flow where no \MDGrevise{linear} TS instability exists.} On increasing $\Wi$, the VNTSA develops multilayered striations of polymer stretch before joining up with 2D EIT through an unstable branch indicated by the red-dashed line in Figure \ref{fig:JFM_2020}a.


\MDGrevise{In the present work, we elaborate on the relationship between  nonlinear Tollmien-Schlichting structures and elastoinertial turbulence.}
In section \ref{sec:coil-stretch}, we describe the evolution of the VNTSA as $\Rey$ increases from $3000$ to $10000$. We detail observations of a coil-stretch transition, with a jump in polymer stretching -- almost two orders of magnitude -- when the local $\Wi$ at the hyperbolic stagnation point of the Kelvin cat's eye structure crosses $\frac{1}{2}$. Accompanying this transition is a structural change from localized striations to extended sheets. These form the necessary ingredients for a process we coin the ``sheet shedding process" responsible for generating multilayered sheets of polymer stretch,  a key characteristic of EIT. We describe this process \MDGrevise{at $\Rey=10000$} and trace its origins in section \ref{sec:TS_to_EIT}.  \MDGrevise{At this Reynolds number, the Newtonian nonlinear Tollmien-Schlichting solution is directly connected to EIT as $\Wi$ increases.} \MDGrevise{Interesting intermittent dynamics occur along this branch, as described in Section \ref{sec:shilnikov}.} \AS{In Section \ref{sec:other_params}, we show the robustness of these observations \MDGrevise{with respect to polymer extensibility and domain size.}}  Finally, in section \ref{sec:broader-context} we \MDGrevise{discuss} the broader context of our results.

\section{Formulation}

This study focuses on two-dimensional pressure-driven channel flow
with constant mass flux. The $x$ and $y$ axes are aligned with the streamwise and wall-normal directions, respectively. 
Lengths are scaled by the half channel height $l$, so the dimensionless channel height $L_y=2$. The domain is periodic in $x$ with length $L_x$. 
Velocity $\mbf{v}$ is scaled with the Newtonian laminar centerline velocity $U$; time $t$ with $l/U$, and pressure $p$ with $\rho U^2$, where $\rho$ is the fluid density. The polymer stress tensor $\mbf{\tau}_p$ is related to the polymer conformation tensor $\mbf{\alpha}$ 
 through the FENE-P constitutive relation, which models each polymer molecule as a pair of beads connected by a nonlinear spring with maximum extensibility $b$.
 
We solve the momentum, continuity and FENE-P equations:

\begin{align}
        \label{Eq_ns_momentum}
                \frac{\partial \mbf{v}}{\partial t} +
                \mbf{v} \cdot \mbf{\nabla v} = -
                \mbf{\nabla}p + \frac{\beta}{Re} \nabla^{2}\mbf{v} +
                \frac{\left(1 -\beta\right)}{Re \Wi_{}}\left(\mbf{\nabla} \cdot
                \mbf{\tau}_{\mathrm{p}}\right), \\
                  \mbf{\nabla} \cdot \mbf{v} = 0,   \\
        	        \mbf{\tau}_p = \frac{\mbf{\alpha}}{1-\frac{\mathrm{tr}(\mbf{\alpha})}{b}} - \mbf{I}, \\             
	    			\frac{\partial \mbf{\alpha}}{\partial t} +         
        			\mbf{v} \cdot \mbf{\nabla \alpha} -
        			\mbf{\alpha} \cdot \mbf{\nabla v} - 
        			\left( \mbf{\alpha} \cdot \mbf{\nabla v} \right)^{\mathrm{T}}
			    = \frac{-1}{\Wi_{}} \mbf{\tau}_p.
\end{align}
Here $Re = \rho U l / (\eta_{\mathrm{s}} + \eta_{\mathrm{p}})$, where
$\eta_s$ and $\eta_p$ are the solvent and polymer contributions to the zero-shear rate viscosity.
The viscosity ratio $\beta = \eta_{\mathrm{s}} / (\eta_{\mathrm{s}} + \eta_{\mathrm{p}})$.  We fix $\beta=0.97$ and $b=6400$ \AS{for all results except in the last section where we use $b = 10^5$}. Since $1-\beta$ is proportional to polymer concentration and $b$ to the number of monomer units, these parameters  correspond to a dilute solution of a high molecular weight polymer.
The Weissenberg number $\Wi = \lambda U/l$, where $\lambda$ is the polymer relaxation time. 

For the nonlinear direct {numerical} simulations {(DNS)} described below, a finite difference scheme and a fractional time step method were adopted for integrating the \MDGreviseII{momentum} equation. \MDGreviseII{Second-order central differences were used for spatial discretization, and second-order Adams-Bashforth and Crank-Nicolson methods were used for time-integration of the convection and diffusion terms, respectively.} The FENE-P equation was discretized using a high resolution central difference scheme, \MDGreviseII{the second-order Kurganov-Tadmor scheme} \citep{kurganov2000new,Vaithianathan:2006dy, Dallas:2010gu}, and, as in \cite{Dallas:2010gu}, time integrated with forward Euler. No artificial diffusion was applied.  \MDGreviseII{A time step of $0.001$ was used, corresponding to a Courant number $\sim 10^{-2}$. Further details and validation cases can be found in \cite{Kushwaha:2017ur} and \cite{Wang:2017tf}.} 
Resolution tests were performed to ensure convergence of statistics. A typical resolution for the following results is $(N_{x},N_{y})$ = ${(131, 602)}$.  This resolution used was based on mesh convergence results
at the highest $\Rey$ and $Wi$ ($10000$ and $13$) considered here. When the resolution was increased to $(N_{x},N_{y})$ = ${(181, 702)}$, the mean
polymer stretch deviations from the laminar base state change by less than 1$\%$. We also perform computations in the shift-reflect symmetric subspace innate to the TS solution family. These computations are done by \AS{simulating half the domain and} applying the following symmetry to generate the fields in the other half:  
	\begin{equation}
	\begin{bmatrix}u & v& \alpha_{xx}& \alpha_{yx}& \alpha_{yy}& \alpha_{zz}\end{bmatrix}(x+L_x/2,y,t)=\begin{bmatrix}u & -v& \alpha_{xx}& -\alpha_{yx}& \alpha_{yy}& \alpha_{zz}\end{bmatrix}(x,-y,t).\label{eq:shiftreflect}
	\end{equation}

\section{Results and discussion}

\subsection{Viscoelastic Tollmien-Schlichting attractor: from $\Rey=3000$ to $10000$ \label{sec:coil-stretch}}

\begin{figure}
	\begin{center}
		\begin{subfigure}{0.45\textwidth}
			\includegraphics[scale=0.4]{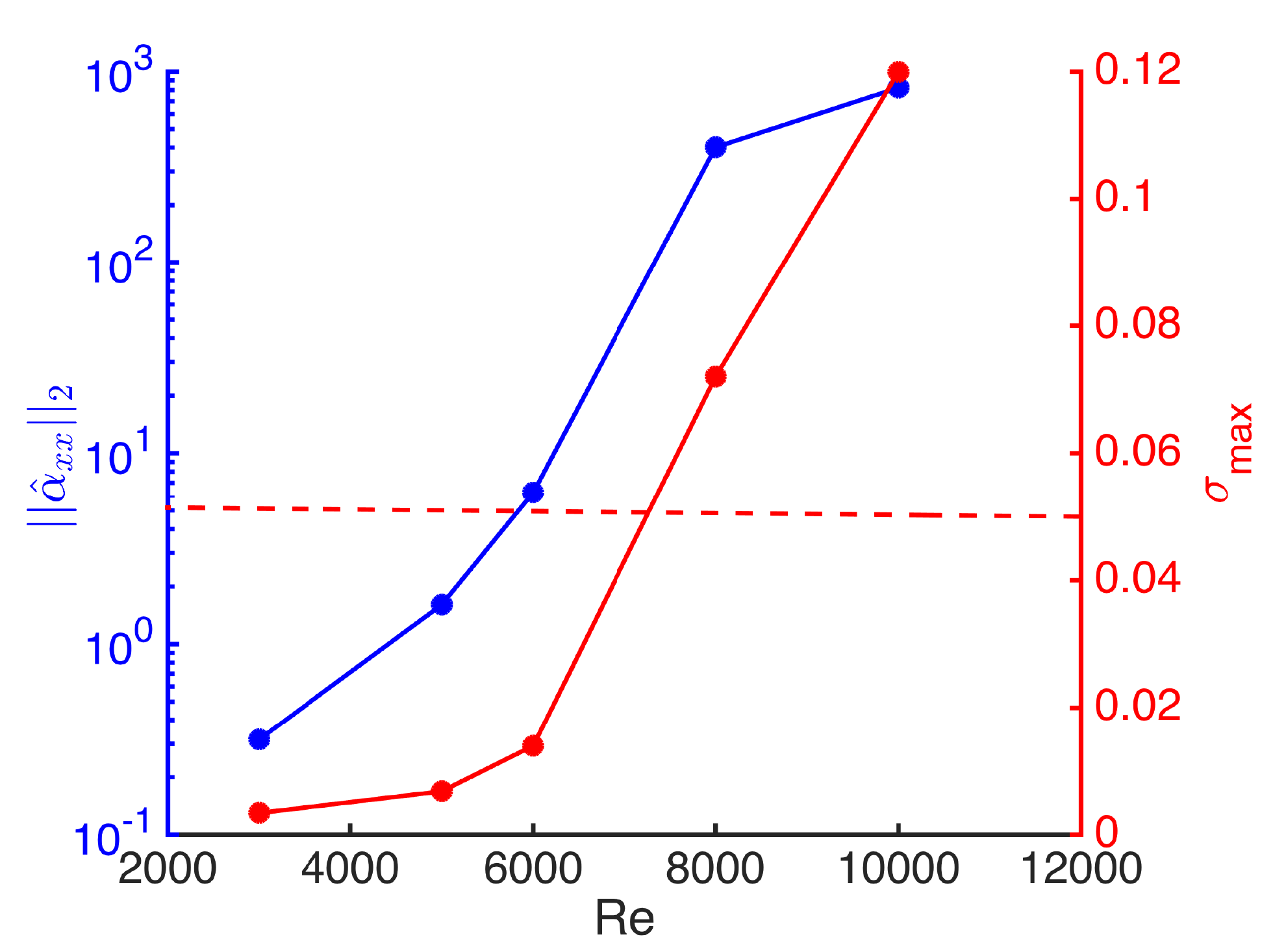} 
			\caption[]{}
			\label{fig:L2_cxxp_vs_Re_Wi10}
		\end{subfigure}\hspace{0.3\textwidth}
		\begin{subfigure}{0.45\textwidth}
	     \includegraphics[scale=0.36]{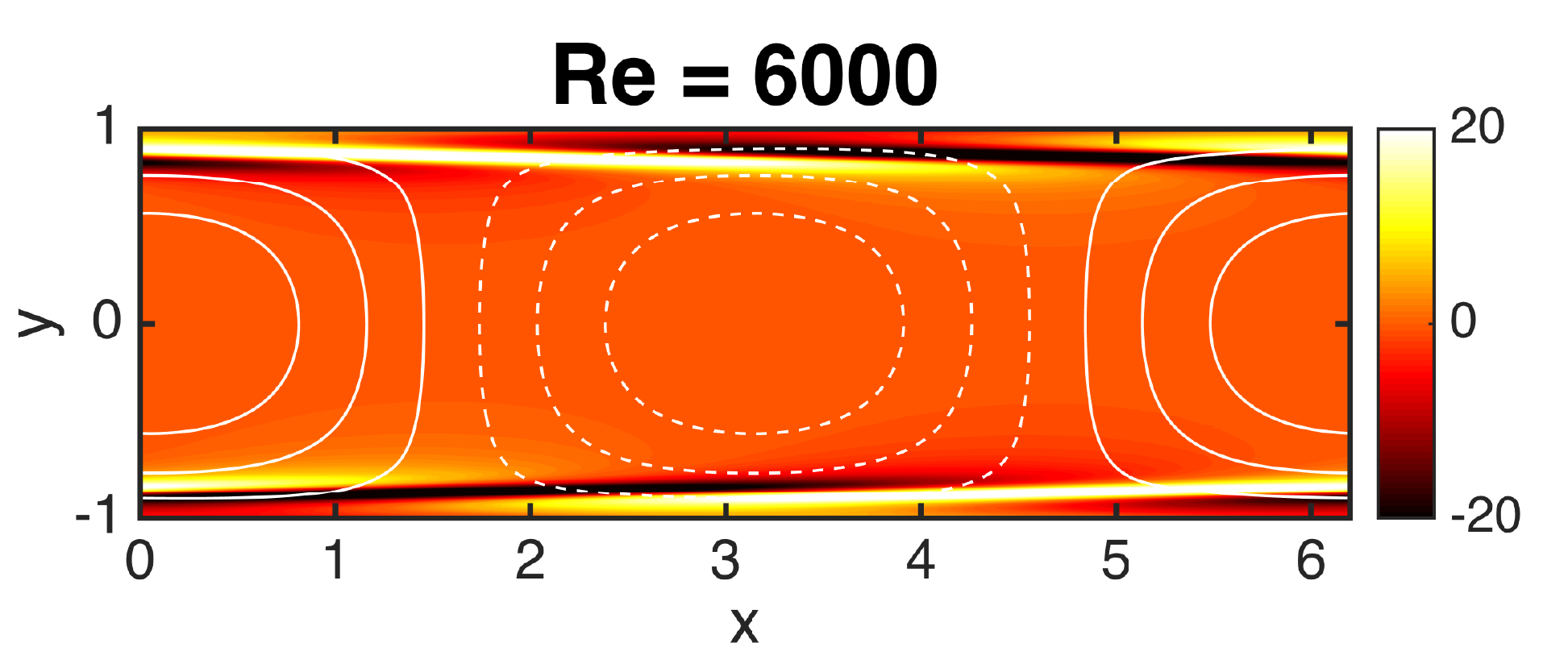} 
      	\label{fig:Re6000_Wi10}
      	\caption[]{}
       \end{subfigure}
		\begin{subfigure}{0.45\textwidth}
	\includegraphics[scale=0.36]{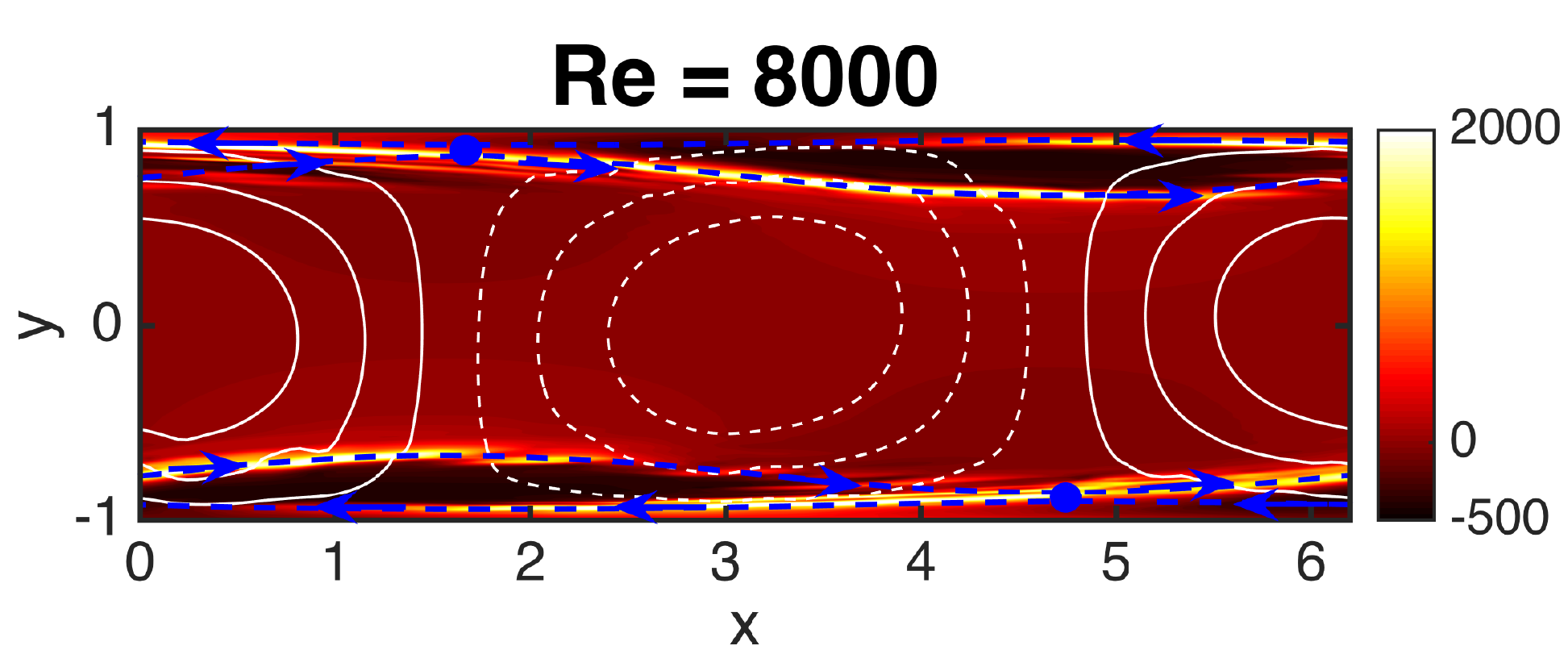} 
	\caption[]{}
	\label{Re8000_Wi10}
		\end{subfigure} 
\end{center}
\caption[]{(a) $L_2$-norm of $\hat{\alpha}_{xx}$ (blue) and corresponding \AS{mean} local stretch rate at the hyperbolic stagnation point (red) along the VNTSA branch at $\Wi = 10$. \AS{Red-dashed line corresponds to $\Wi_\mathrm{loc} = \lambda \sigma_\mathrm{max} = 1/2$}. (b) and (c) are snapshots of the fluctuation structure of the solution branch at $Re = 6000$ and $8000$ respectively. Shown are contour lines of $\hat{v}$ superimposed on color contours of $\hat{\alpha}_{xx}$. Here $\hat{}$ denotes deviations from laminar base state. \MDGreviseII{On (c), the hyperbolic stagnation points in the traveling frame are indicated as blue dots, and the streamlines attached to them as dashed curves with arrows indicating the direction of flow.}}
\label{fig:Coil_stretch}	     
\end{figure} 

\AS{The results presented in this section make the connection between the results of \MDGrevise{Shekar et al.~\cite{Shekar:2020gt}} and the bulk of the present work, which will focus on the attractor at $\Rey = 10000$.} We begin by considering the evolution of the  VNTSA as $\Rey$ is increased from $3000$ to $10000$ at fixed $\Wi = 10$, $L_x = 6.2$. \MDGreviseII{To find these solutions, an initial condition of the laminar state perturbed by the linear TS mode was used to land on the attractor at $\Rey = 3000$, following which the velocity and polymer stretch fields from the nonlinear TS attractor at a one $\Rey$ were used as initial conditions at a higher $\Rey$ to trace out the evolution of the attractor from $\Rey=3000$ to $10000$. We then performed the same process but moving downward in $\Rey$ from the solution at $10000$ to verify that there is no hysteresis on this solution branch upon changing $\Rey$. }
 \AS{Note the difference in box size from the results in \cite{Shekar:2020gt}.} \MDGrevise{In the earlier study, $L_x$ was chosen based on the lowest $\Rey$ for existence of finite amplitude nonlinear TS waves. In contrast, in the present work, $L_x$ was chosen to approximately correspond to the domain size at which the linear TS instability first appears (at $\Rey=5772$) \citep{DrazinReid}. This was done so that we could examine, as we do below, the behavior when the TS mode is linearly unstable to explore the relationship between the NNTSA, VNTSA and EIT. Indeed, at the value $L_x=5$ chosen in the earlier work, the TS mode does not become linearly unstable at any $\Rey$.   }

The blue curve on Figure \ref{fig:Coil_stretch}a quantifies \MDGrevise{the evolution of the VNTSA} in $\Rey$ using the \MDGrevise{time-averaged} $L_2$-norm of $\hat{\alpha}_{xx}$ as a measure. \AS{Unless otherwise noted, all quantities reported here are time-averaged.} With the attractor being weakly chaotic, at least 3000 \MDGrevise{time units} (TU) of data, and in some cases up to 6000 TU,  is used to get an accurate estimate of this quantity. \MDGrevise{The key observation is that in the range  $6000<\Rey<8000$, $||\hat{\alpha}_{xx}||_2$ increases by several orders of magnitude.} \MDGrevise{Note that} to capture this large change, a log scale is used. Figures \ref{fig:Coil_stretch}b and c are representative snapshots of the flow and polymer conformation fields, which illustrate the structural change accompanying this transition. \AS{Throughout this manuscript, we report the deviations from the laminar state, rather than the total. Otherwise, for cases where the TS structure is weak, especially in the case of $\balpha$, it would not be discernible.} 

\MDGrevise{As at $\Rey=3000$,} the attractor at $\Rey = 6000$ (figure \ref{fig:Coil_stretch}b) continues to display striations of weak polymer stretch fluctuations that are localized near the wall, and closely resembles the linear TS mode.  Apart from the obvious quantitative differences, the attractor at 8000 (figure \ref{fig:Coil_stretch}c) now displays a sheet of \MDGrevise{very strong} polymer stretch \AS{emanating from the near-wall stagnation point regions}. 
 \MDGrevise{Additionally, at $\Rey=8000$ we see that, in the region near the bottom wall where the wall-normal velocity is negative (roughly $1.6<x<4.6$), the sheet of high stretching seems to have broken into several smaller sheets, and similarly near the top wall where $v_y>0$. We elaborate below on the ``sheet shedding" process that leads to this structure, and how it connects the TS structure to EIT.} 
 
To shed insights into the nature of this transition, we examine the flow kinematics more quantitatively. Aside from a very weak temporally nonperiodic component, the flow is a traveling wave displaying the characteristic Kelvin cat's eye streamlines, and the corresponding hyperbolic stagnation points, in the traveling frame. Denoting the largest eigenvalue of the velocity gradient $\nabla\mathbf{v}^T$ at these points as $\sigma_\mathrm{max}$, \MDGreviseII{material lines near these points will stretch along the unstable direction of the stagnation point as $\exp(\sigma_\mathrm{max}t)$. Polymer molecules near these points will also stretch exponentially if $\lambda\sigma_\mathrm{max}>1/2$ \citep{de1974coil,Graham:2018ty}.}  Defining a local Weissenberg number $\Wi_\mathrm{loc}=\lambda\sigma_\mathrm{max}$, this criterion becomes $\Wi_\mathrm{loc}>1/2$. \MDGreviseII{
Figure \ref{fig:Coil_stretch}a reports estimates (red) of $\sigma_\mathrm{max}$ vs.~$\Rey$. Values are in the range $0.01\lesssim \sigma_\mathrm{max}\lesssim 0.1$ (in units of $U/l$), indicating that $\Wi_\mathrm{loc}\ll \Wi$. The origin of this difference in magnitude can be understood as follows: recall that the laminar velocity field is everywhere a pure shear flow, for which the straining and vortical (symmetric and antisymmetric) components of $\nabla\mathbf{v}^T$ are equal, both of its eigenvalues are zero, it has only one eigenvector, $[1,0]^T$, and material line stretching is everywhere linear in time, no matter how large $\Wi$ is. The velocity field of the VNTSA is a relatively weak perturbation away from the laminar flow, and accordingly, the eigenvalues of the velocity gradient tensor are weakly perturbed from zero: $\sigma_\mathrm{max}\ll 1$ or equivalently $\Wi_\mathrm{loc}\ll\Wi$. Furthermore, at the hyberbolic stagnation points, the eigenvectors of $\nabla\mathbf{v}^T$ (the unstable and stable directions) are nearly aligned, because they both emerge from the single eigenvector $[1,0]^T$ of the degenerate pure shear case.  Figure \ref{fig:Coil_stretch}c shows the positions of the hyperbolic stagnation points (blue dots) and the streamlines emanating from these points at $\Rey=8000$. Observe the small angle between the streamlines entering and leaving the stagnation points, as well as the sheet of highly stretched fluid oriented along the unstable direction. At $\Rey=6000$ the angle is too small to see clearly so we opted not to show the streamlines for that case.   }

Since the attractors are not pure traveling waves, the estimates of $\sigma_\mathrm{max}$ in Figure \ref{fig:Coil_stretch}a are based on time-averaging over multiple snapshots sampled over the attractor. 
\MDGreviseII{For a given snapshot, say at $t=2000$, the fields at $t = 1999$ and $t =2001$ would be used to estimate the instantaneous wavespeed by calculating the streamwise distance the position of the maximum wall normal velocity moves.}
This wavespeed is then subtracted off of the streamwise velocity to yield the velocity field in the traveling frame, from which the stagnation point positions can be identified and $\sigma_\mathrm{max}$ estimated.  The drastic change  between $\Rey = 6000$ and $8000$ coincides with $\Wi_\mathrm{loc}$ crossing the coil-stretch threshold $\Wi_\mathrm{loc}=1/2$.

\begin{figure}
	\begin{center}
		\begin{subfigure}{0.55\textwidth}
			\includegraphics[scale=0.35]{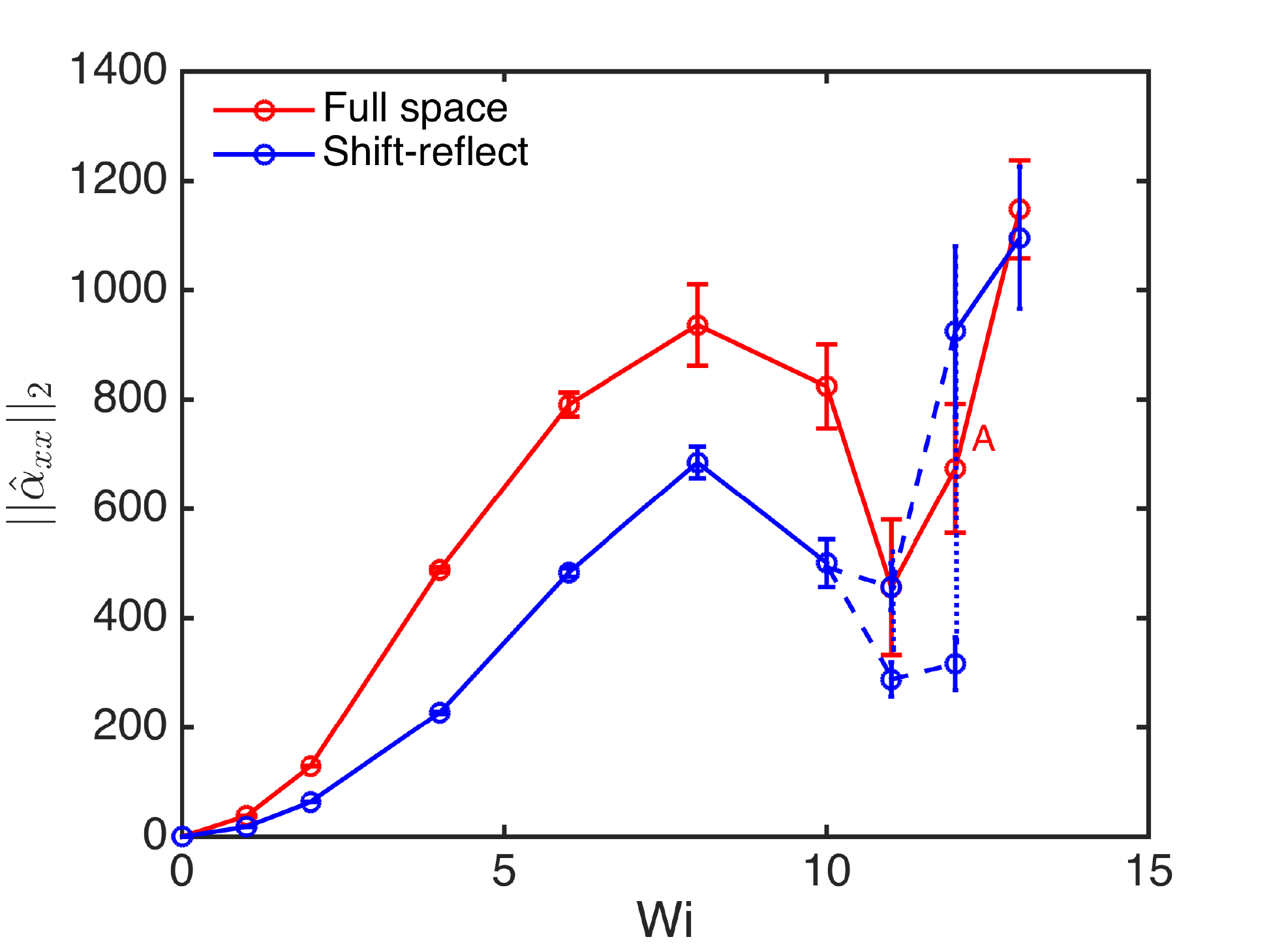}
			\caption[]{}
			\label{fig:L2_Cxxp_vs_Wi_FS}
		\end{subfigure}
		\hspace{0.3\textwidth}
		\vspace*{-0.18in}
		\begin{subfigure}{0.45\textwidth}
	     	\includegraphics[scale=0.3]{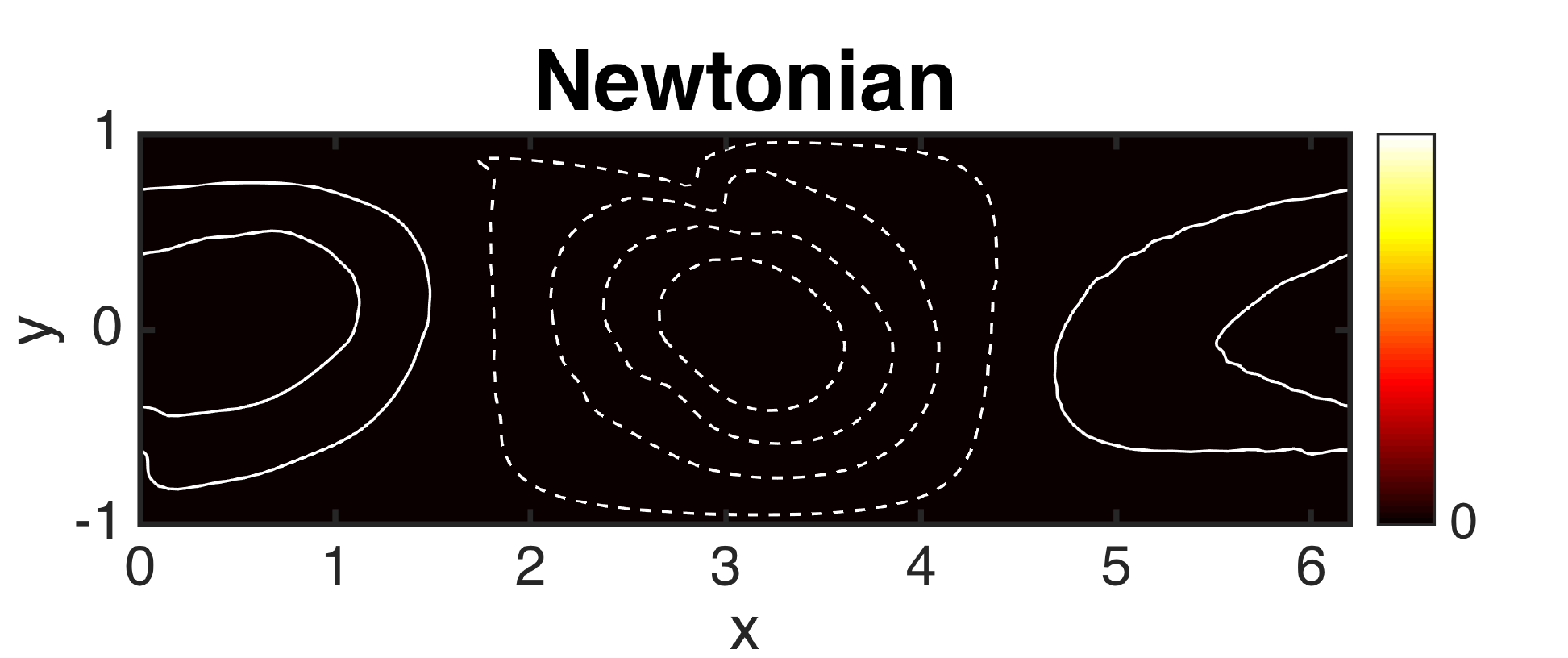} 
	     	\caption[]{}
      		\label{fig:Re10000_Newt_FS}
        \end{subfigure}
        \begin{subfigure}{0.45\textwidth}
	     	\includegraphics[scale=0.3]{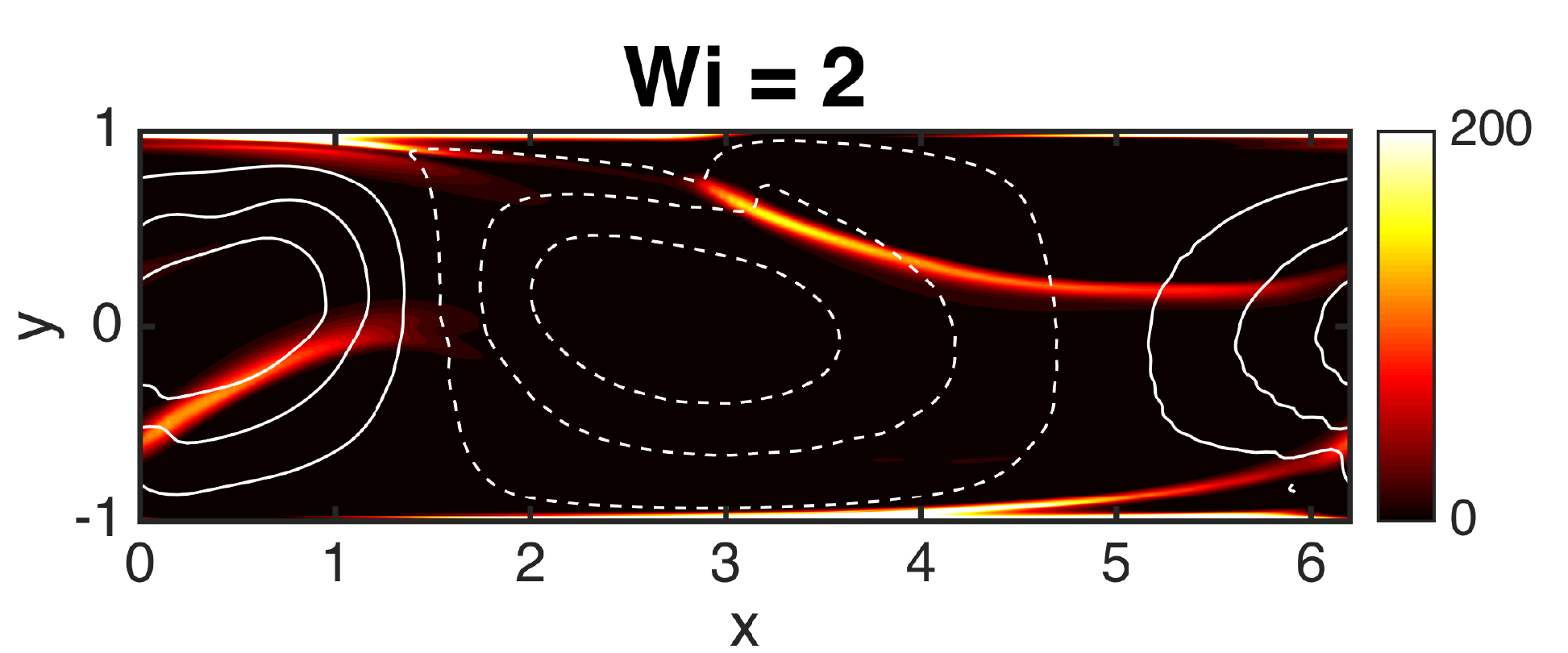} 
	     	\caption[]{}
      		\label{fig:Combined_3980_Wi2}
        \end{subfigure}
  		\vspace*{-0.11in}      
        \begin{subfigure}{0.45\textwidth}
	     	\includegraphics[scale=0.3]{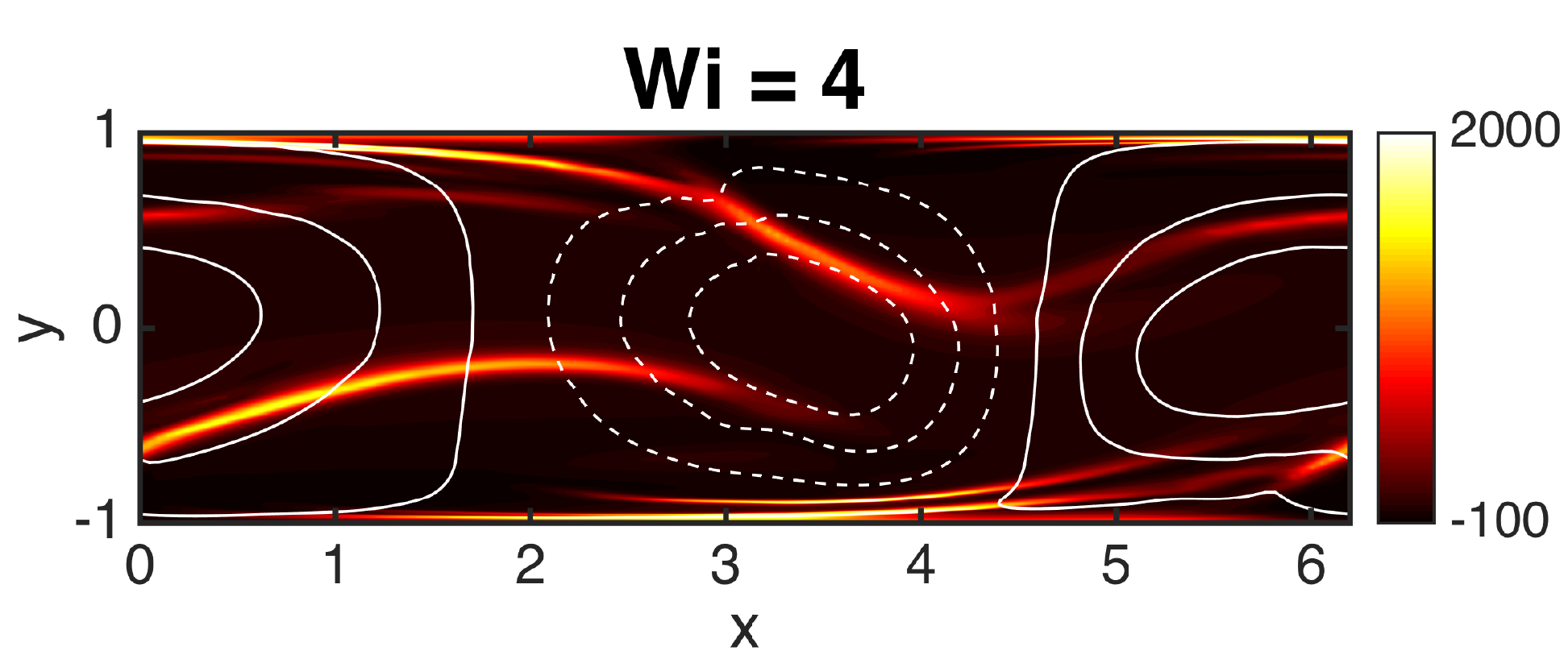} 
	     	\caption[]{}
      		\label{fig:TSA_Wi4}
        \end{subfigure}
        \begin{subfigure}{0.45\textwidth}
	     	\includegraphics[scale=0.3]{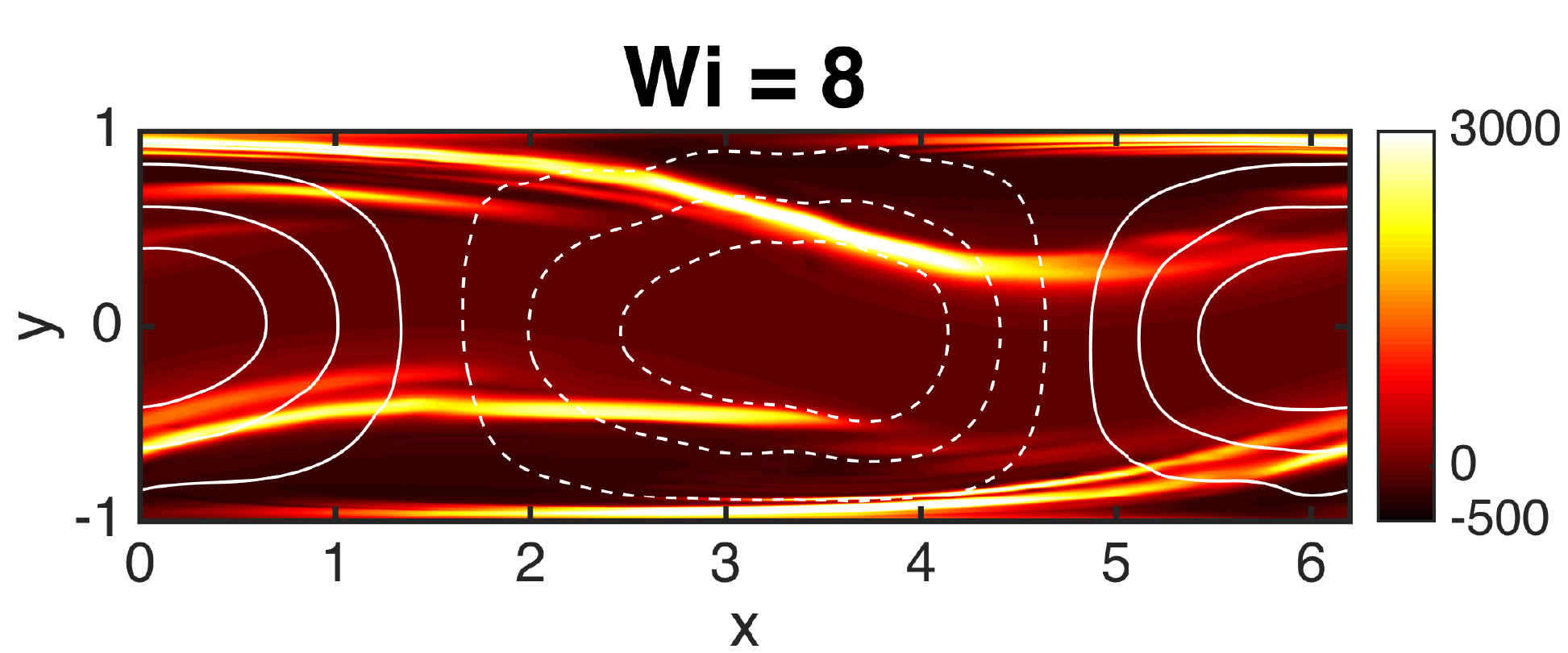} 
	     	\caption[]{}
      		\label{fig:TSA_Wi8}
        \end{subfigure}
         		\vspace*{-0.11in}
       \begin{subfigure}{0.45\textwidth}
	     	\includegraphics[scale=0.3]{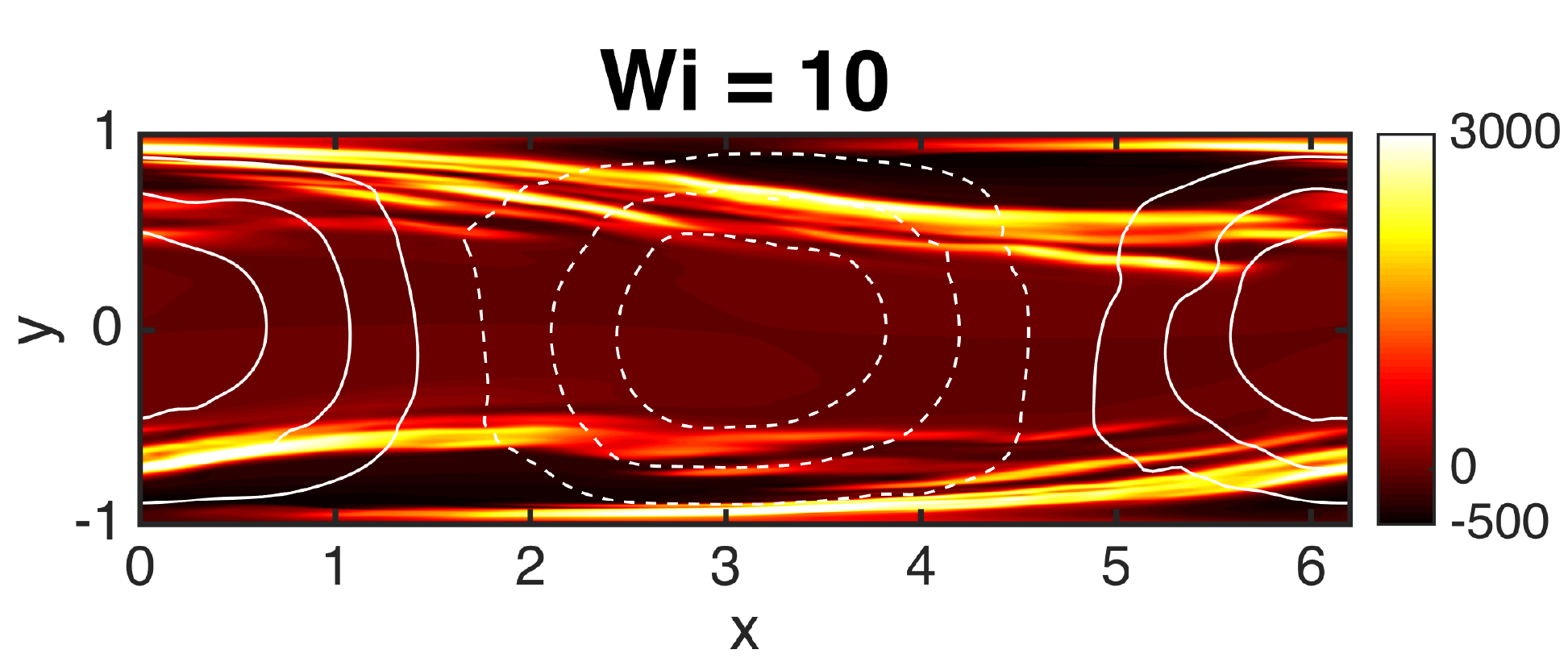} 
	     	\caption[]{}
      		\label{fig:TSA_Wi10}
        \end{subfigure}
        \begin{subfigure}{0.45\textwidth}
	     	\includegraphics[scale=0.3]{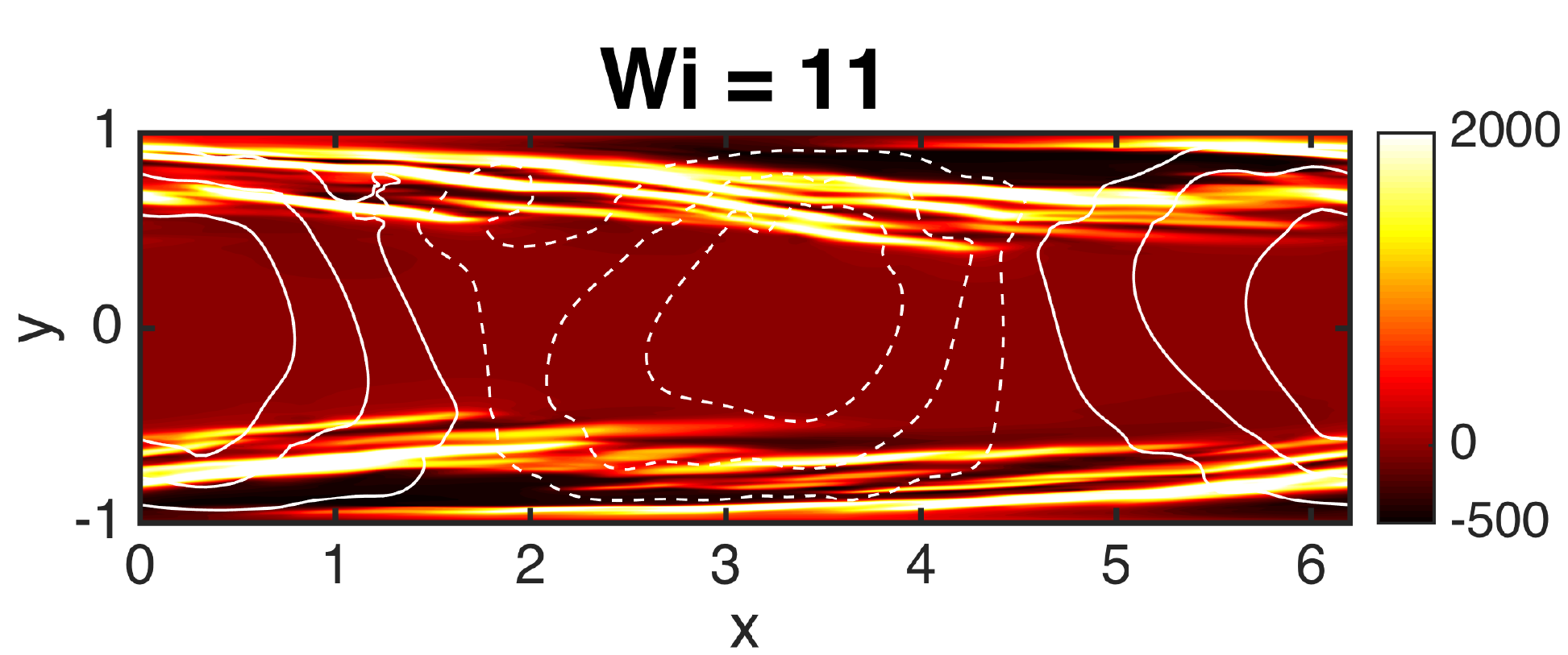} 
	     	\caption[]{}
      		\label{fig:TSA_Wi11}
        \end{subfigure}
        		\vspace*{-0.11in}
		\begin{subfigure}{0.45\textwidth}
			\includegraphics[scale=0.3]{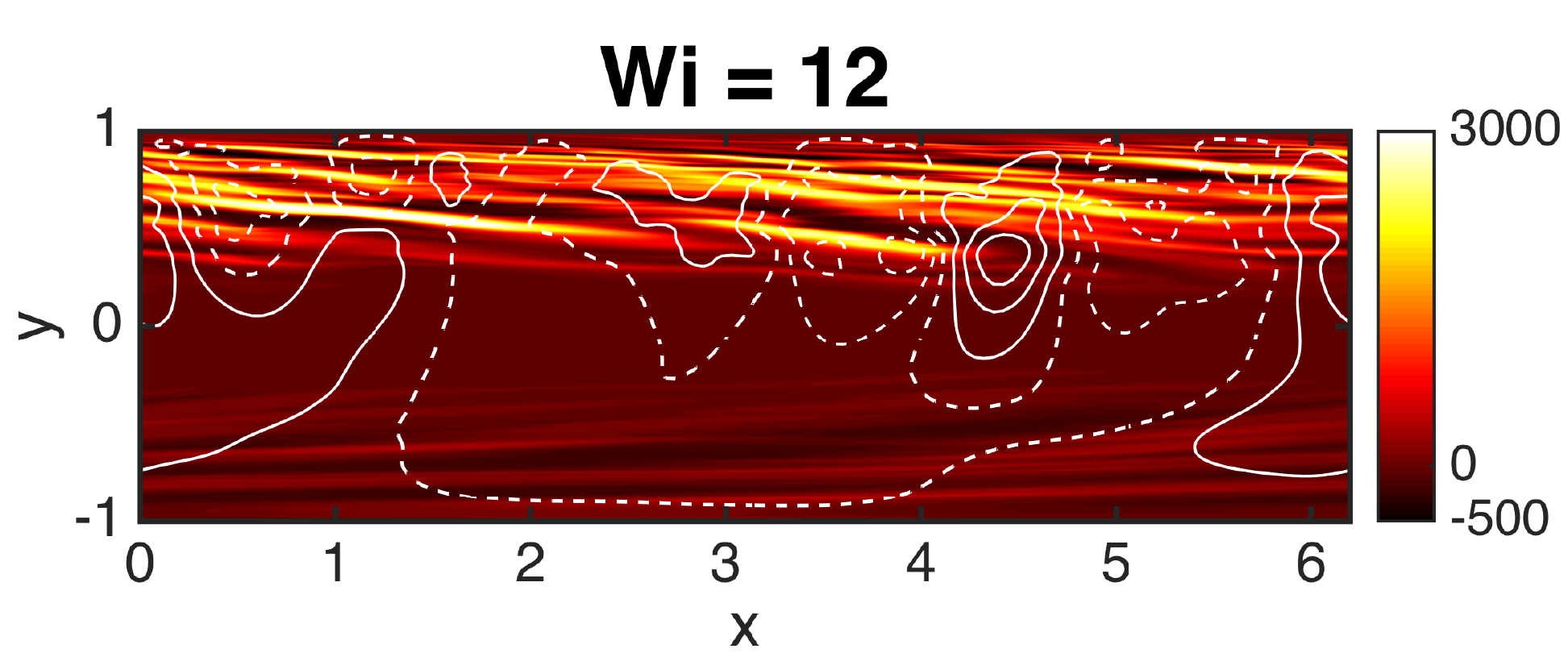} 
      		\label{fig:Re10000_Wi12_FS}
      		\caption[]{}
       \end{subfigure}
		\begin{subfigure}{0.45\textwidth}
			\includegraphics[scale=0.3]{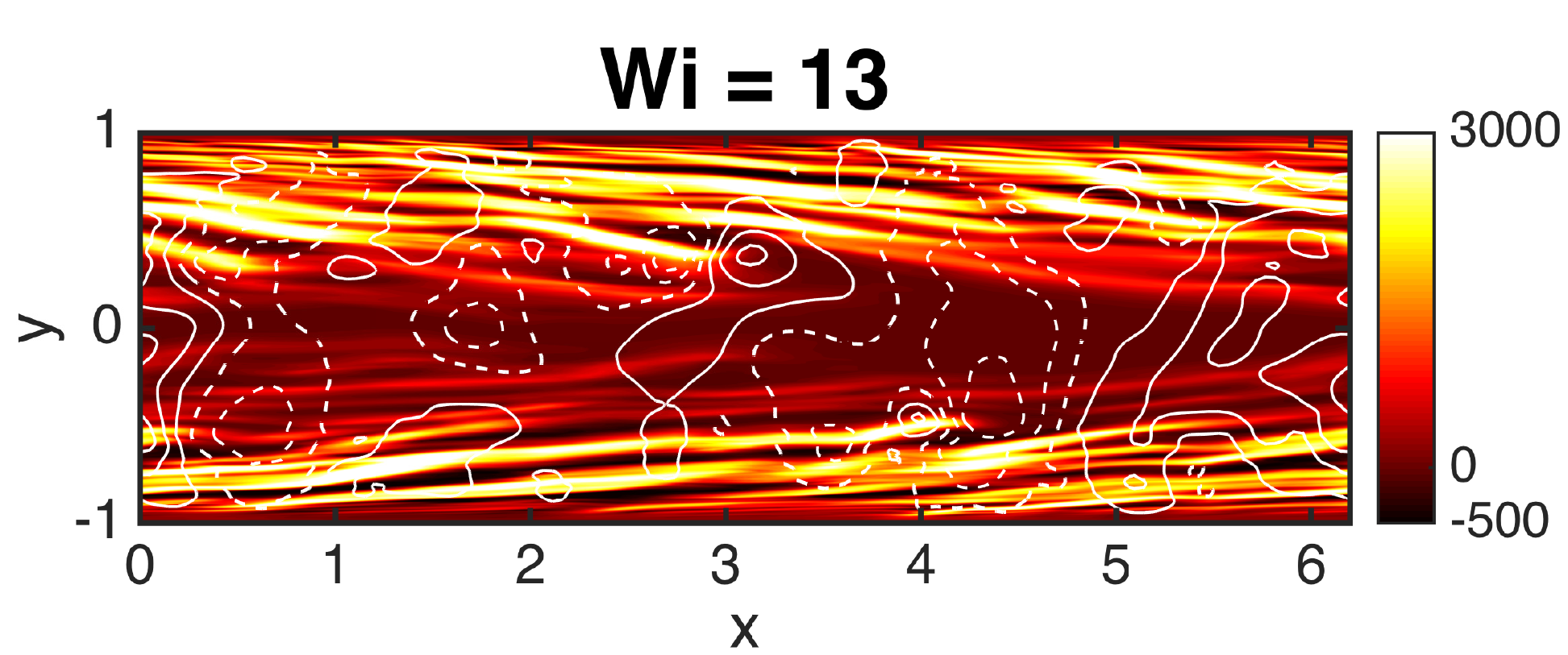} 
			\caption[]{}
			\label{fig:Re10000_Wi13_FS}
		\end{subfigure}
		        		\vspace*{-0.15in}

\end{center}
\caption[]{(a) $||\hat{\alpha}_{xx}||_2$ vs $\Wi$ at $\Rey = 10000$ \MDGrevise{ for simulations in the full space and the shift-reflect subspace}. Here, `A' indicates asymmetry in the attractor in the full space. \MDGrevise{The upper and lower blue symbols at $\Wi=11$ and $12$ indicate averages over the two metastable states intermittently visited by the dynamics at these $\Wi$ values.}  (b) - (i) are snapshots of the fluctuation structure at the indicated $\Wi$. }
\label{fig:Re10000_bif_dia_FS}	     
\end{figure}

\subsection{$\Rey=10000$: Sheet-shedding and the evolution from TS waves to EIT\label{sec:TS_to_EIT}}

\MDGrevise{At $\Rey=3000$, the solution branch described in the previous section loses existence as $\Wi$ is decreased to zero.  In contrast, at $\Rey=10000$, }
 \AS{the attractor presented in the previous section at $\Wi = 10$ can be traced continuously and non-hysteretically back to the Newtonian attractor on decreasing $\Wi$.} \MDGrevise{Thus, at these conditions, the NNTSA and VNTSA are no longer distinct solution branches, so we simply refer to the TS attractor. Equally importantly, the solution at $\Wi=10$ evolves, with a slight further increase in $\Wi$, to EIT}. \MDGrevise{\emph{That is, at $\Rey=10000$, the Newtonian nonlinear Tollmien-Schlichting solution evolves continuously to EIT as $\Wi$ increases.}} 
 
\begin{figure}
	\begin{center}
		\begin{subfigure}{0.45\textwidth}
			\includegraphics[scale=0.36]{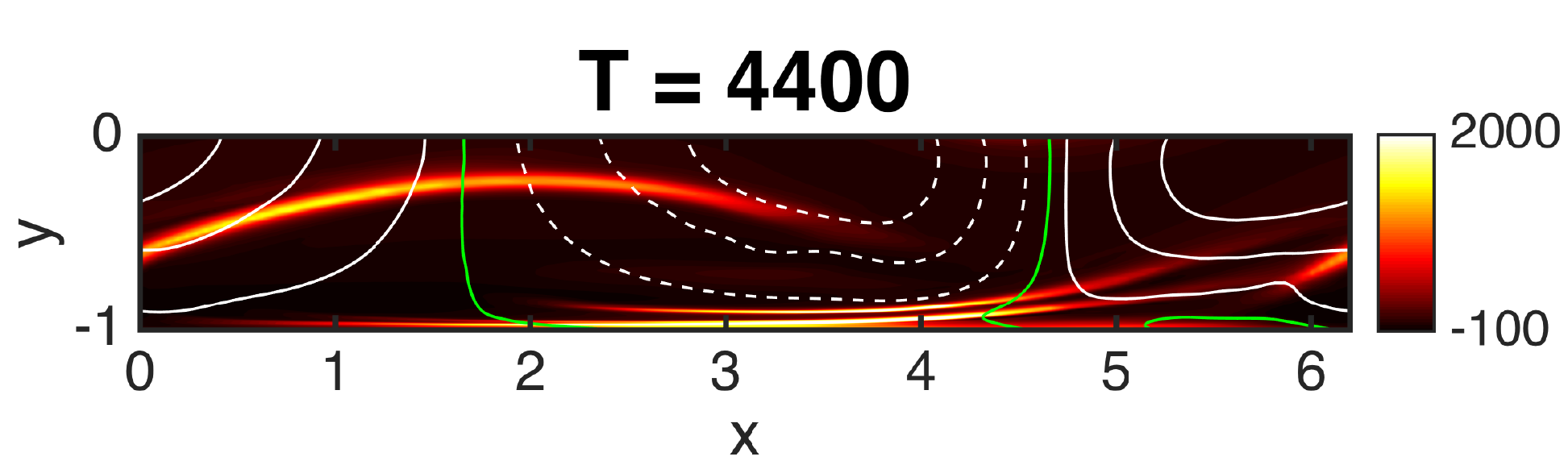} 
			\caption[]{}
			\label{fig:SSP_Wi4_4400}
		\end{subfigure}
		\begin{subfigure}{0.45\textwidth}
			\includegraphics[scale=0.36]{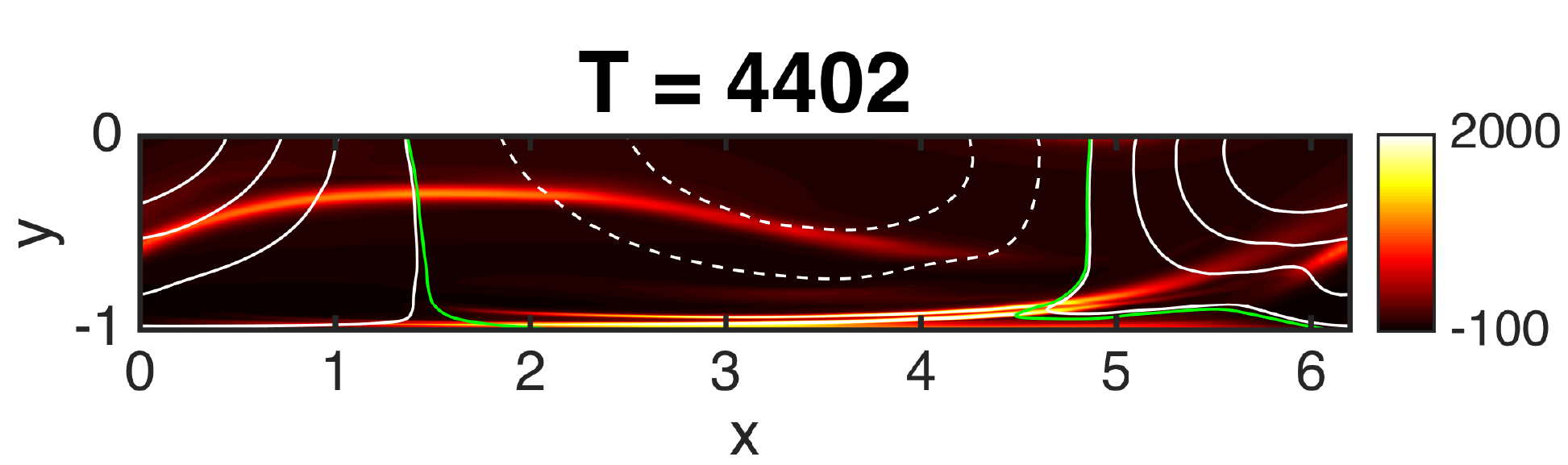} 
			\caption[]{}
			\label{fig:SSP_Wi4_4402}
		\end{subfigure}
		\begin{subfigure}{0.45\textwidth}
	    	\includegraphics[scale=0.36]{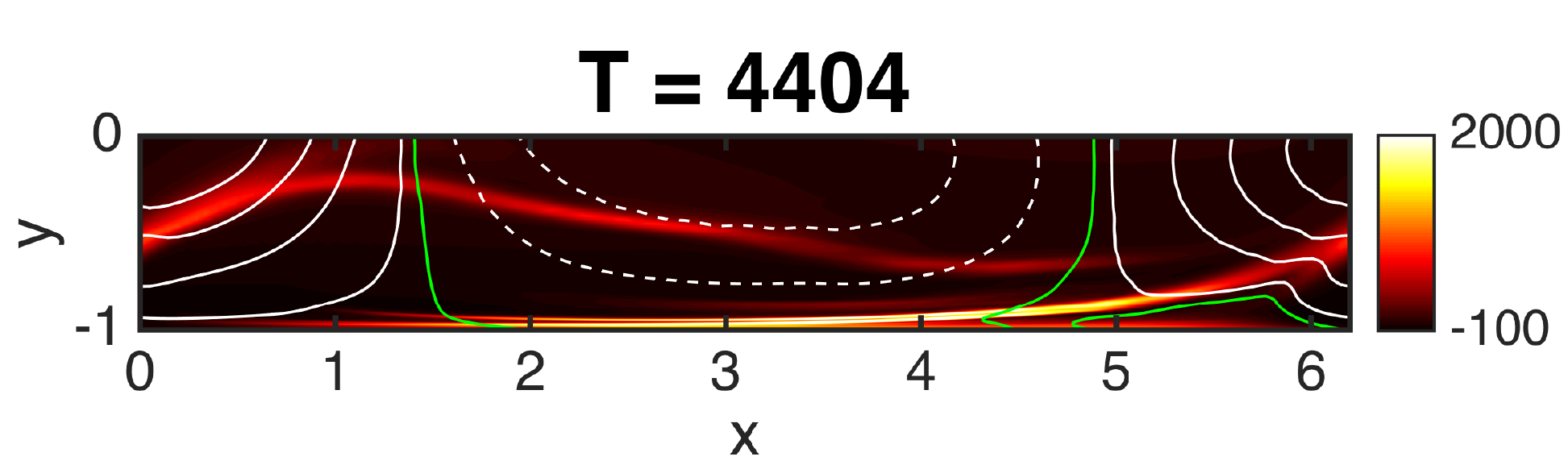}
	    	\caption[]{}
      		\label{fig:SSP_Wi4_4404}
		\end{subfigure}
		\begin{subfigure}{0.45\textwidth}
			\includegraphics[scale=0.36]{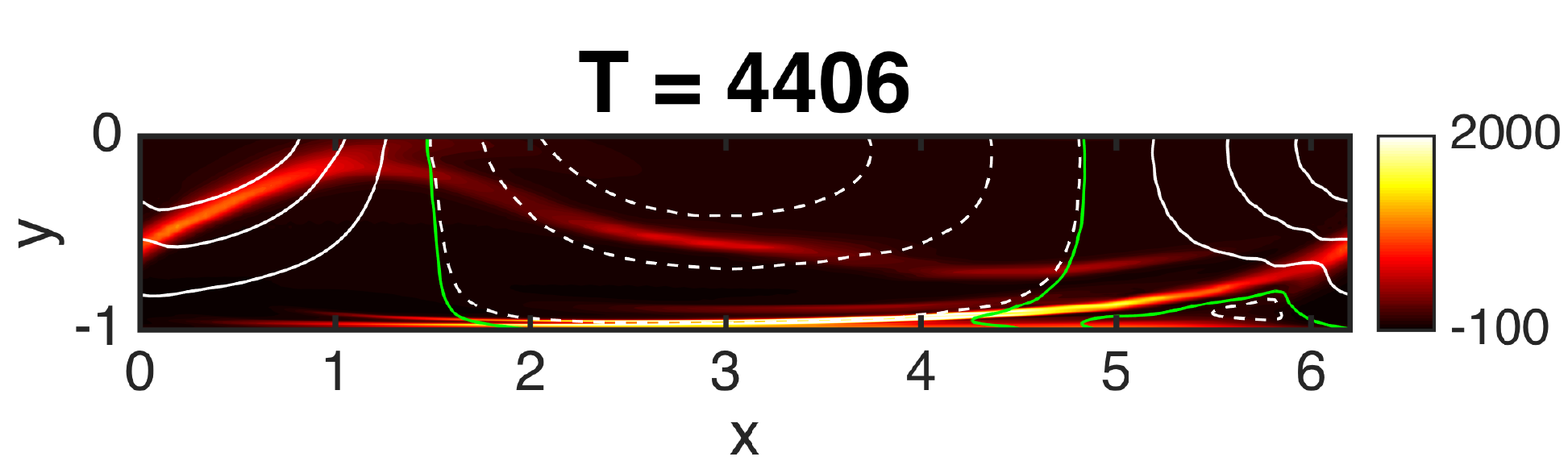}
			\caption[]{} 
			\label{fig:SSP_Wi4_4406}
		\end{subfigure} 
		\begin{subfigure}{0.45\textwidth}
			\includegraphics[scale=0.36]{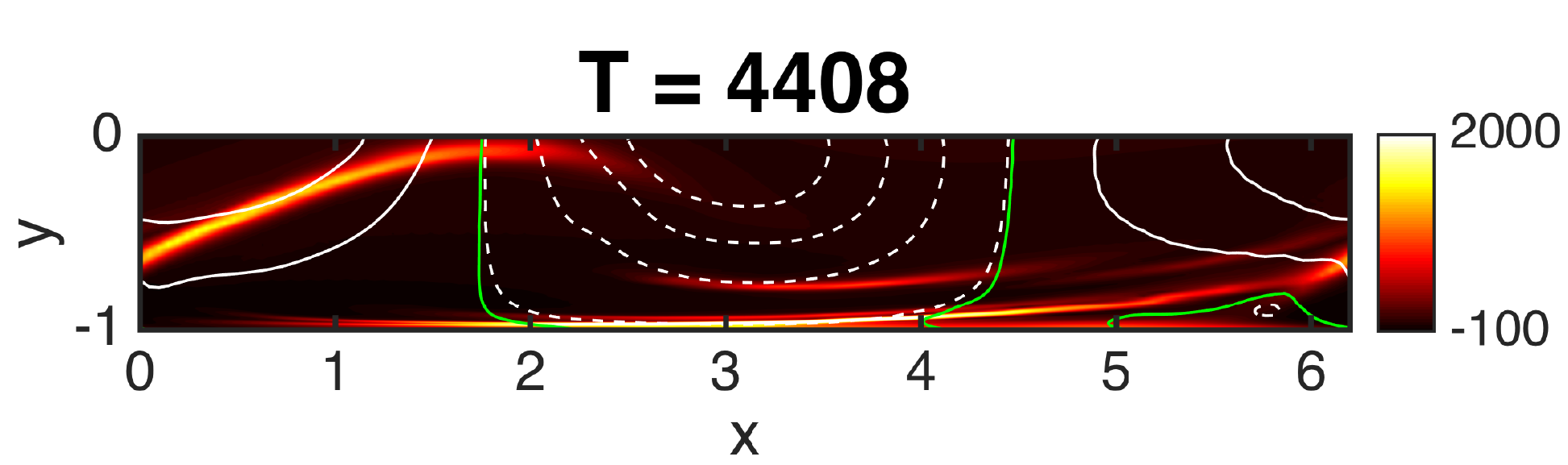}
			\caption[]{} 
      		\label{fig:SSP_Wi4_4408}
       \end{subfigure}
		\begin{subfigure}{0.45\textwidth}
			\includegraphics[scale=0.36]{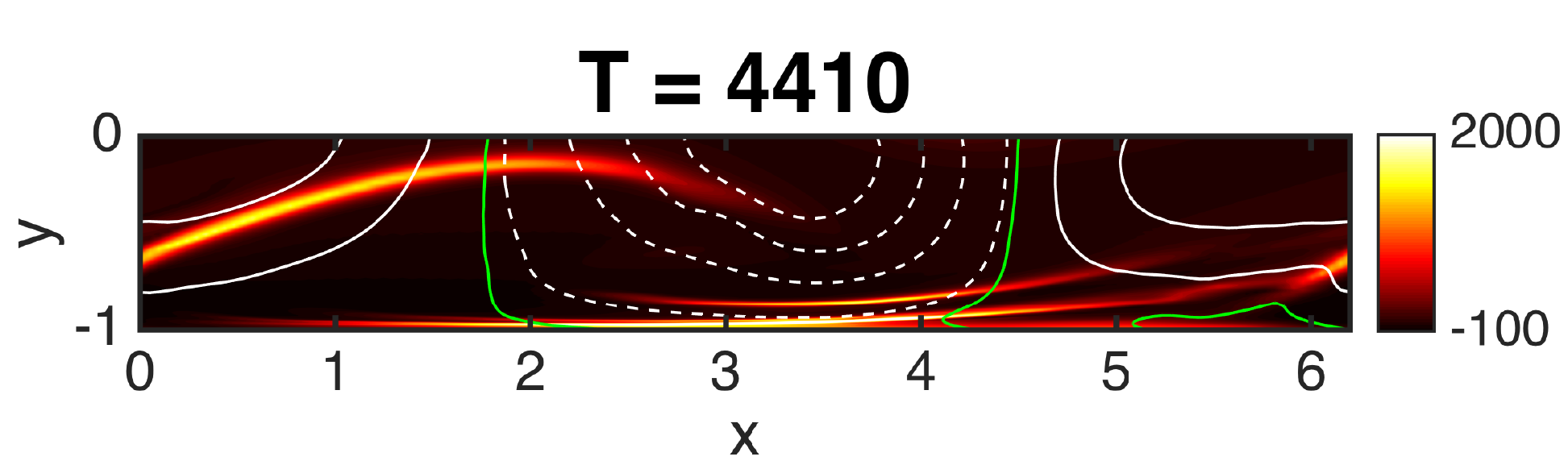} 
			\caption[]{}
			\label{fig:SSP_Wi4_4410}
		\end{subfigure}
\caption[]{(a) - (f):  snapshots of the fluctuation structure at $\Rey = 10000$, $\Wi = 4$. Snapshots are taken from $t = 4400$ to $4410$ every $2$ time units respectively. Green lines correspond to $\hat{v} = 0$, i.e., where wall normal velocity changes sign. Bottom half of the domain shown.}
\label{fig:SSP_Wi4}	     
\end{center}
\end{figure}

\AS{Figure \ref{fig:Re10000_bif_dia_FS}a  shows this evolution using $||\hat{\alpha}_{xx}||_2$ as the measure.} \MDGreviseII{To generate the results in this section, initial conditions of the laminar base state perturbed by the linear TS mode were used to land on the nonlinear attractor at each $\Wi$. Furthermore, a second set of runs were performed using the fields from the nonlinear attractor at a certain $\Wi$ as initial conditions at higher or lower $\Wi$. Results on both increasing and decreasing $\Wi$ were consistent with the first set of runs, indicating the absence of hysteresis. }\AS{Also shown on this plot are the standard deviations of this statistic to give a sense of the temporal intermittency of the dynamics along the branch.} \AS{Here, the results in the full space (red) are shown along with the results in the shift-reflect (S-R) symmetric subspace (blue).}  The full space attractor displays an increasing trend in $||\hat{\alpha}_{xx}||_2$ from Newtonian to $\Wi = 8$ followed by a decrease up to $\Wi = 11$ and an increase beyond that. \AS{The instantaneous dynamics of $||\hat{\alpha}_{xx}||_{2}$ go from  2-torus dynamics up to $\Wi = 2$ to mild chaos at $\Wi = 4$.}  As  $\Wi$ further increases, the dynamics get more complicated and reach a peak in intermittency around $\Wi = 11$. Here, intermittent bursts characterized by a sharp temporary increase in the instantaneous $||\hat{\alpha}_{xx}||_{2}$ can be seen as a persistent phenomena. 

\begin{figure}[t]
	\begin{center}
		\vspace*{-0.25in}
		\begin{subfigure}{0.45\textwidth}
			\includegraphics[scale=0.31]{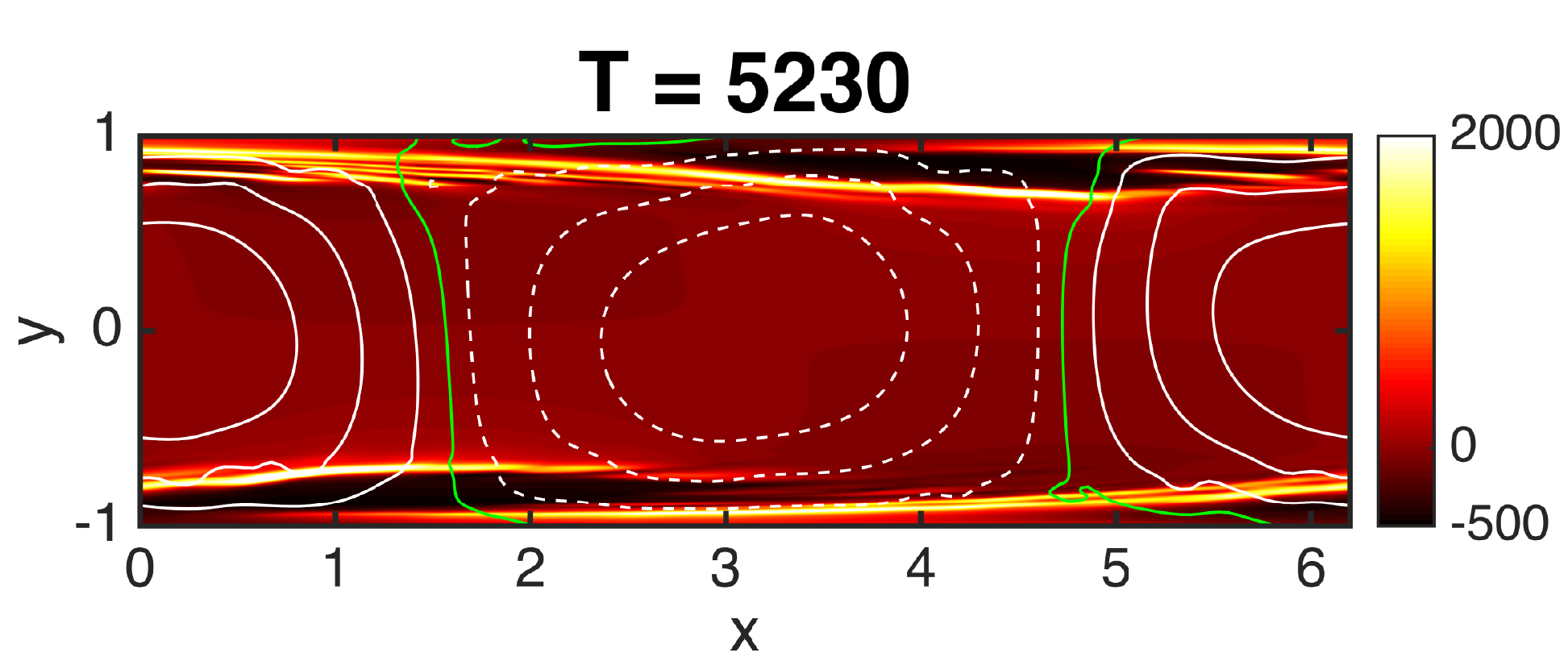} 
			\caption[]{}
			\label{fig:Combined_5230}
		\end{subfigure}
		\begin{subfigure}{0.45\textwidth}
			\includegraphics[scale=0.31]{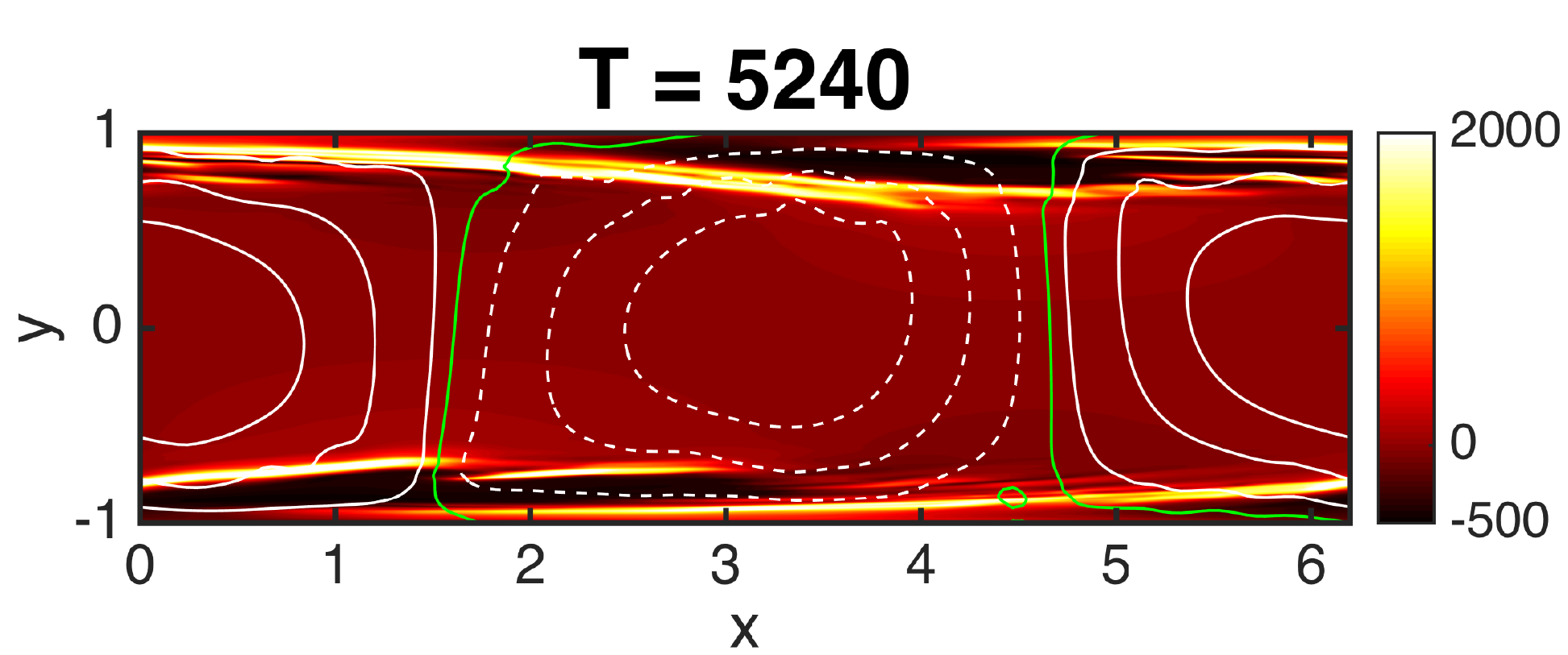} 
			\caption[]{}
			\label{fig:Combined_5240}
		\end{subfigure}
		\begin{subfigure}{0.45\textwidth}
	    	\includegraphics[scale=0.31]{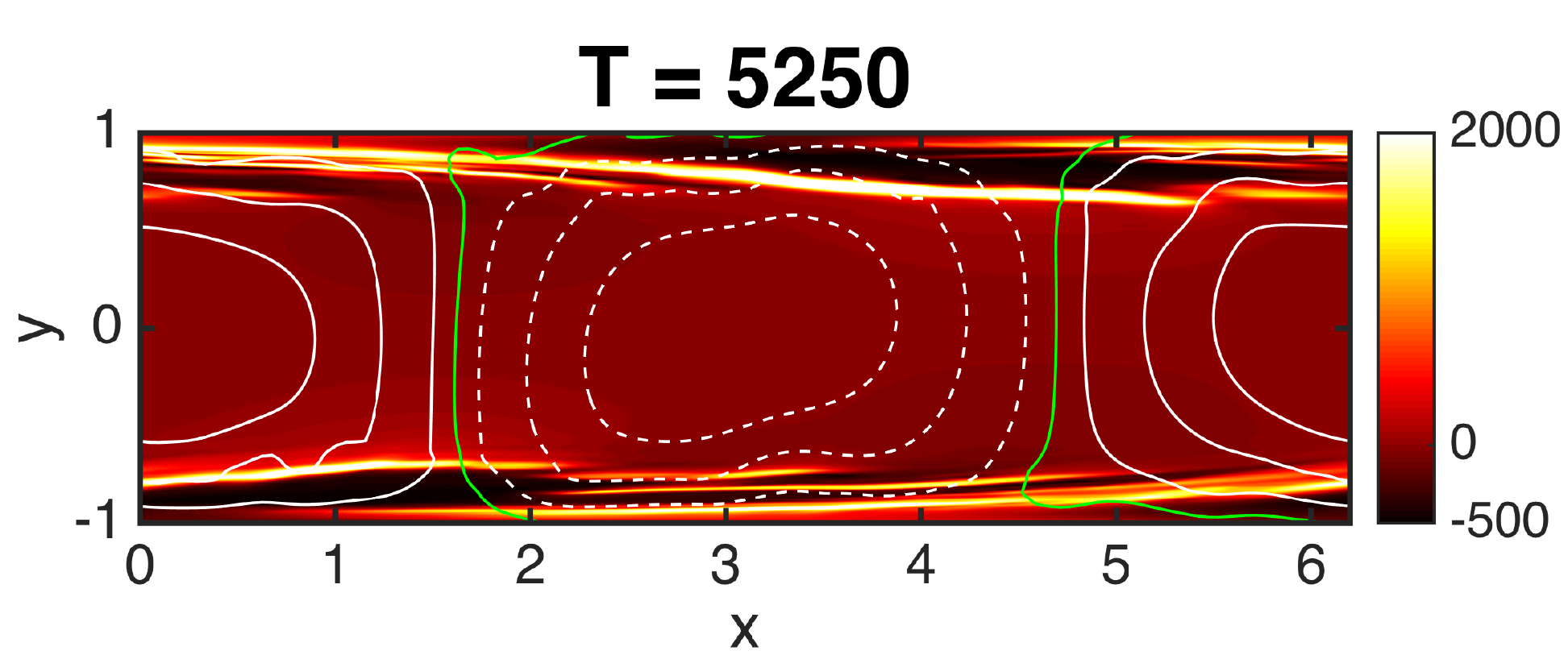}
	    	\caption[]{}
      		\label{fig:Combined_5250}
		\end{subfigure}
		\begin{subfigure}{0.45\textwidth}
			\includegraphics[scale=0.31]{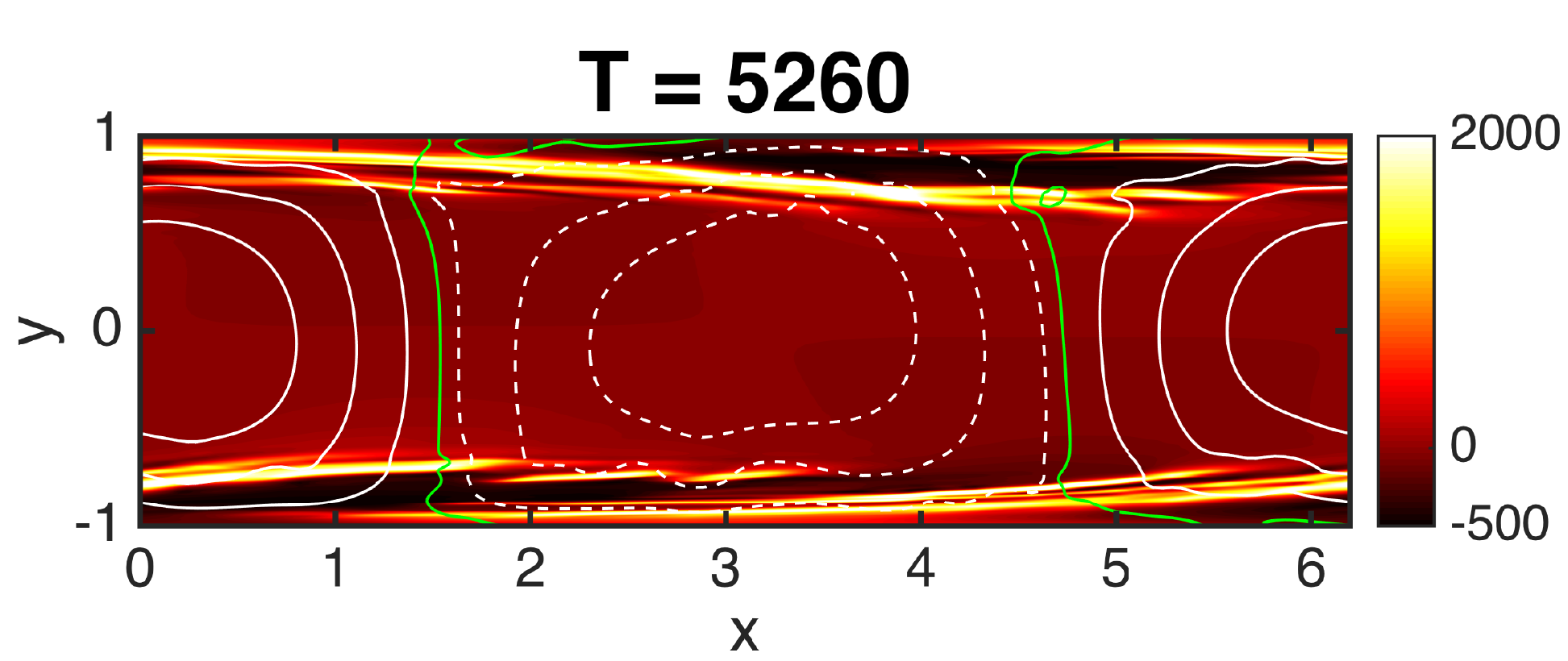}
			\caption[]{} 
			\label{fig:Combined_5260}
		\end{subfigure} 
		\begin{subfigure}{0.45\textwidth}
			\includegraphics[scale=0.31]{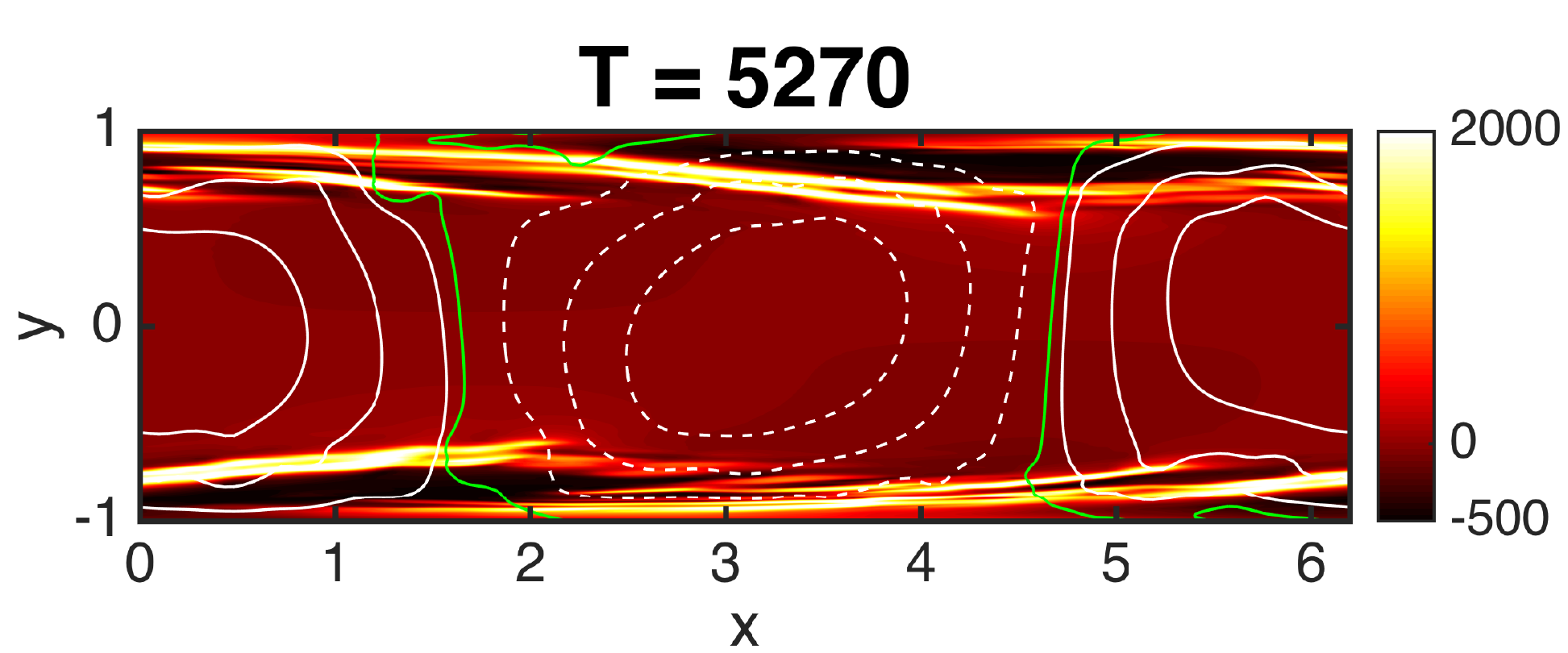}
			\caption[]{} 
      		\label{fig:Combined_5270}
       \end{subfigure}
		\begin{subfigure}{0.45\textwidth}
			\includegraphics[scale=0.31]{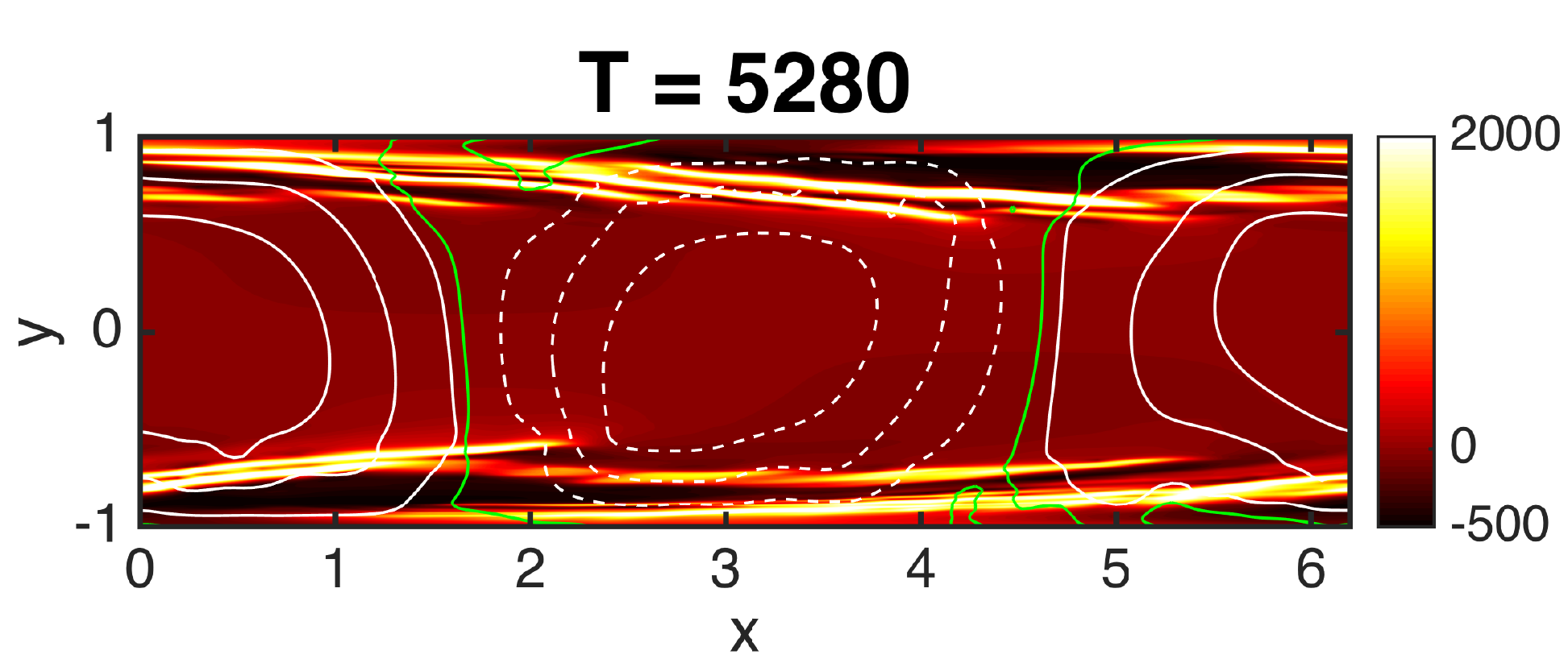} 
			\caption[]{}
			\label{fig:Combined_5280}
		\end{subfigure}
				\begin{subfigure}{0.45\textwidth}
	    	\includegraphics[scale=0.31]{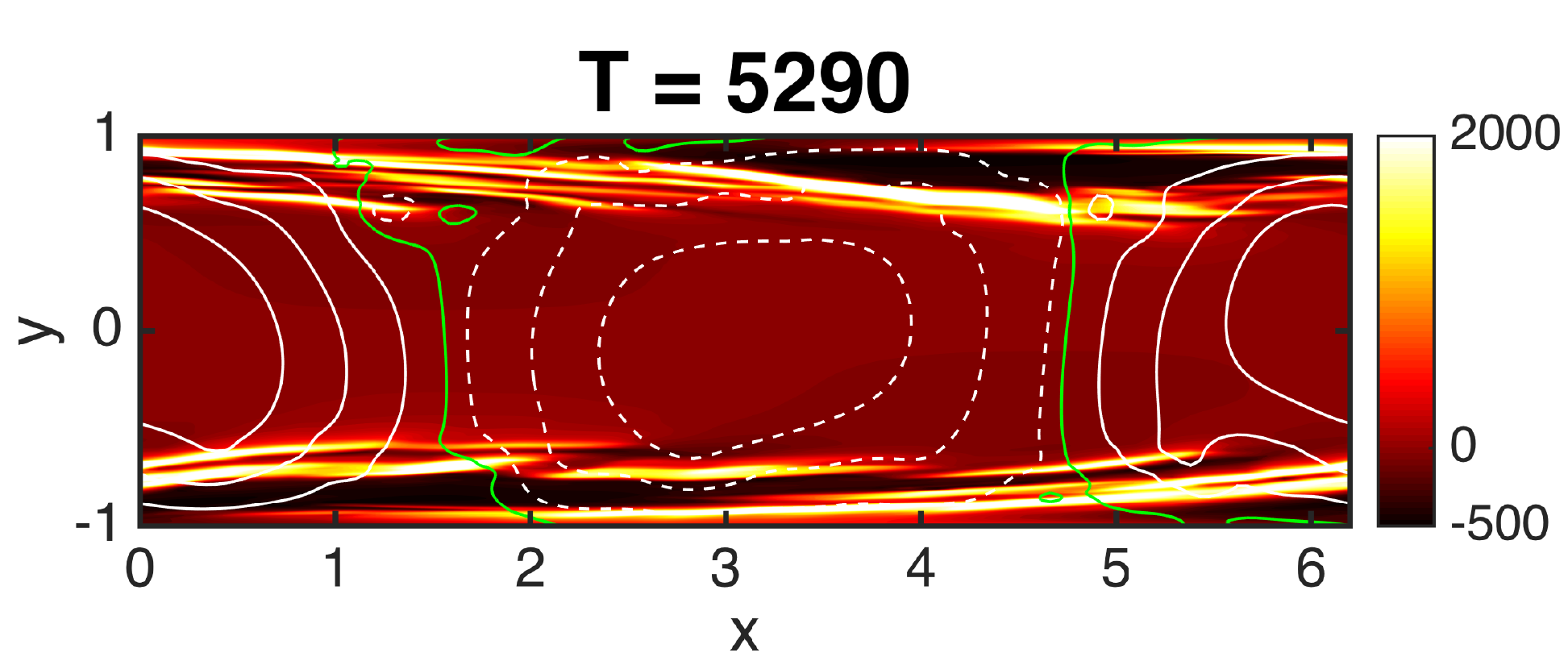}
	    	\caption[]{}
      		\label{fig:Combined_5290}
		\end{subfigure}
		\begin{subfigure}{0.45\textwidth}
			\includegraphics[scale=0.31]{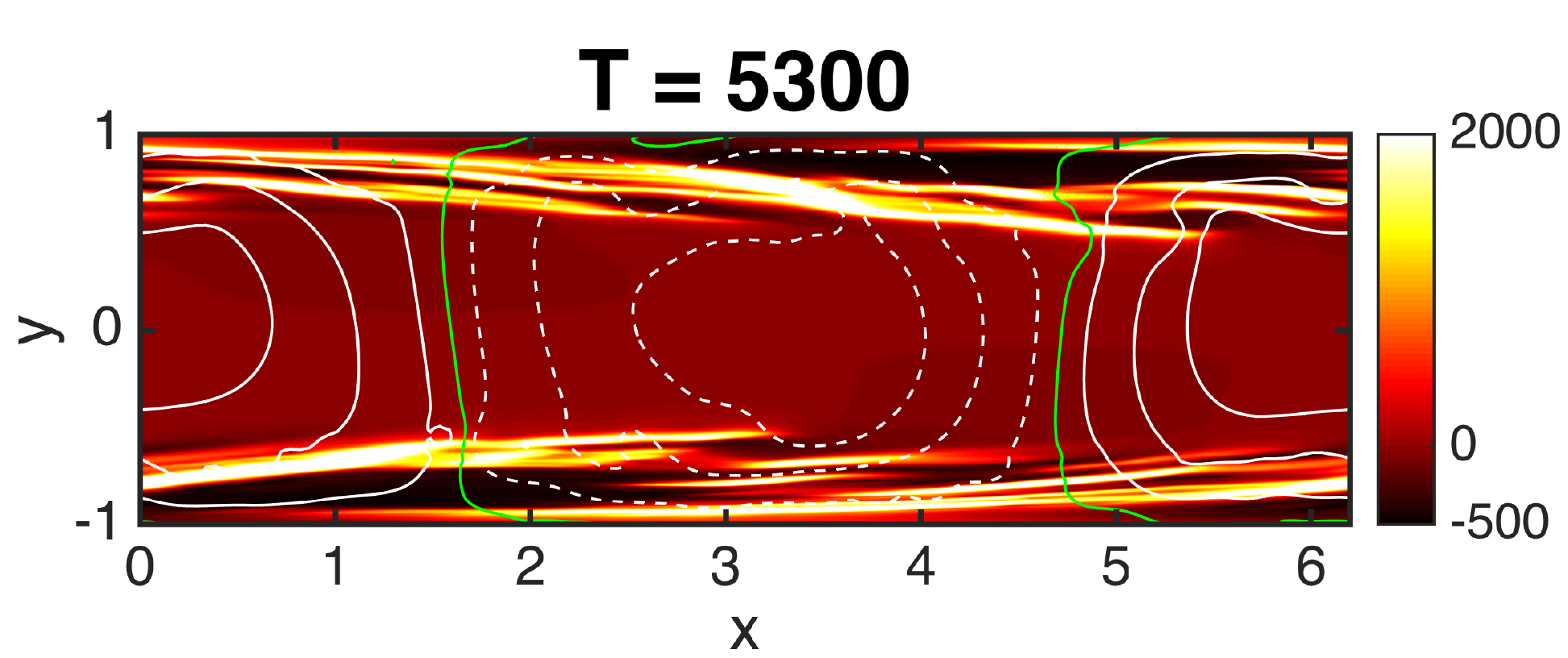}
			\caption[]{} 
			\label{fig:Combined_5300}
		\end{subfigure} 
		\begin{subfigure}{0.45\textwidth}
			\includegraphics[scale=0.31]{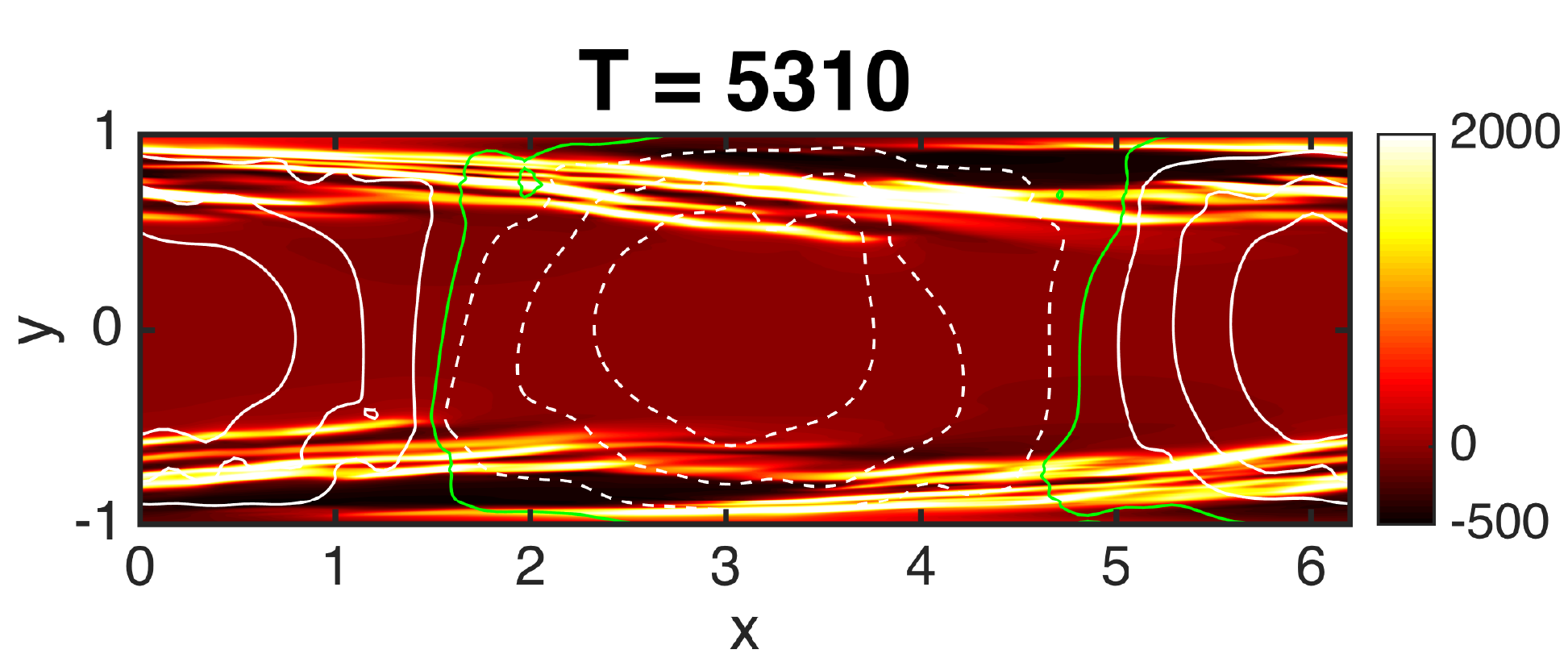}
			\caption[]{} 
      		\label{fig:Combined_5310}
       \end{subfigure}
		\begin{subfigure}{0.45\textwidth}
			\includegraphics[scale=0.31]{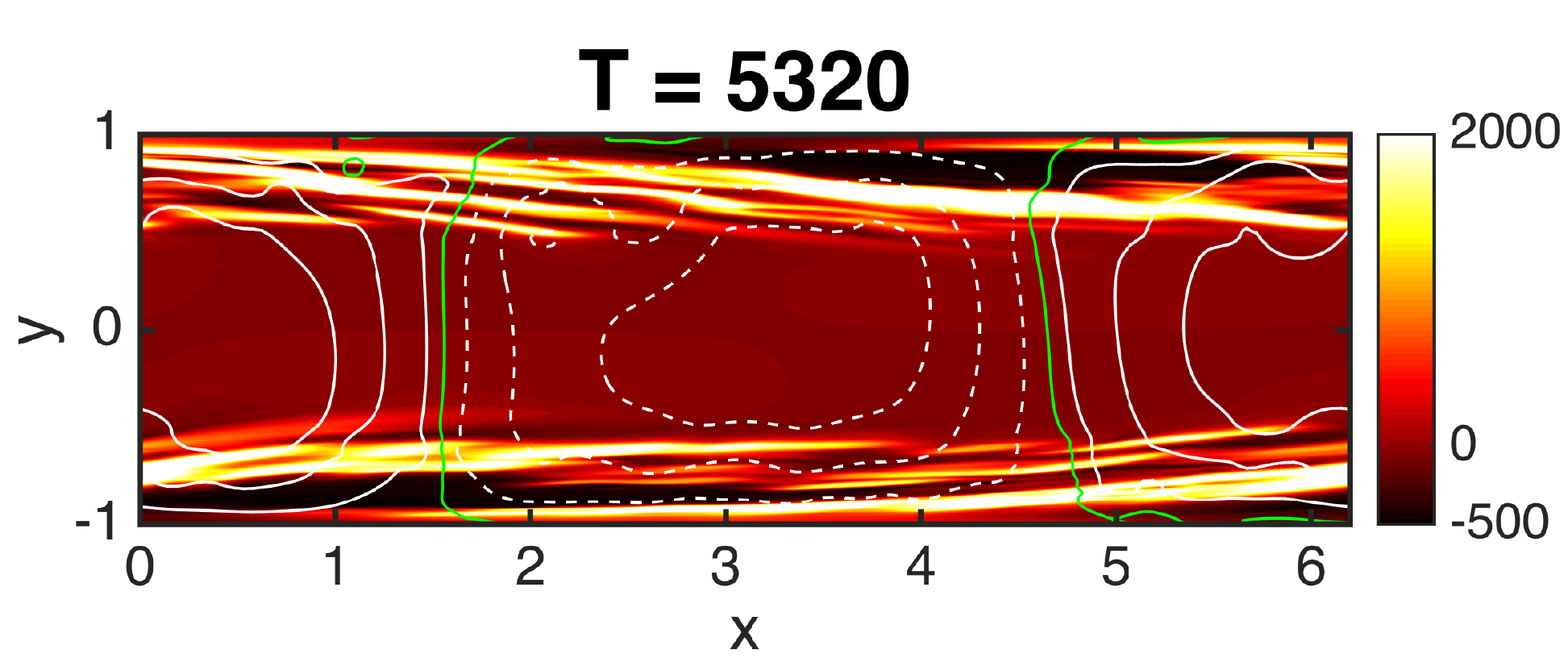}
			\caption[]{} 
			\label{fig:Combined_5320}
		\end{subfigure}
\caption[]{(a) - (j): snapshots of the fluctuation structure at $\Wi = 11$. Snapshots are taken from $t = 5230$ to $5320$ every $10$ time units respectively. Same format as Figure \ref{fig:SSP_Wi4}.}
\label{fig:SSP_Wi11}	     
\end{center}
\end{figure}

\AS{Many of the observations described in the full space carry forward to the S-R subspace up to $\Wi = 10$.} \AS{The same non-monotonic trend in $||\hat{\alpha}_{xx}||_{2}$ is accompanied by a temporal modulation around $\Wi = 4$.} \AS{In other words, the full space attractor can be seen as a modification of the attractor in the S-R subspace.} \AS{Beyond $\Wi = 10$, the dynamics in the S-R space alternate between two metastable states (indicated by vertical dotted lines) before settling on the stable symmetric attractor at $\Wi = 13$.} \AS{We will expand on this observation in Section \ref{sec:shilnikov}. }

To get a sense of what underlies the dynamics, we now turn to the full-space flow structures at various  \MDGrevise{values of $\Wi$. Snapshots of these structures are shown in Figures \ref{fig:Re10000_bif_dia_FS}b-i.}
The Newtonian attractor (figure \ref{fig:Re10000_bif_dia_FS}b) displays the usual wall normal velocity contour lines across the channel height. At  $\Wi = 2$ (figure \ref{fig:Re10000_bif_dia_FS}c), the attractor develops a sheet of polymer stretch that starts near the wall and arches closer to the centerline. This is typical of the NNTSA structure described in the Introduction and can be attributed to the Kevin cat's eye critical layer kinematics of the attractor. \AS{Going from $\Wi = 2$ to $4$, the attractor starts to display near-wall multilayered sheets, as seen between $x \approx 3$ and $x \approx 5$ in Figure \ref{fig:Re10000_bif_dia_FS}d.} \AS{This effect gets more pronounced as we move along the branch, with $\Wi = 11$ intermittently displaying strong multilayered sheets (figure \ref{fig:Re10000_bif_dia_FS}g) similar to those seen at EIT at $\Wi = 13$ (figure \ref{fig:Re10000_bif_dia_FS}i).} Interestingly, $\Wi = 12$ displays strong multilayered sheets on one half of the channel with no particular preference towards either half. (\AS{By symmetry, for every state with fluctuations localized near the top, there is a dynamically equivalent state localized near the bottom.} \AS{Asymmetric states similar to this have been observed in 2D Newtonian channel flow \cite{Kerswell.2021}.}) 
 The marked structural transition beyond $\Wi = 11$ coincides with an increasing trend in $||\hat{\alpha}_{xx}||_{2}$ in Figure \ref{fig:Re10000_bif_dia_FS}a and giving rise to the EIT branch beyond $\Wi = 11$.  \AS{We now describe the process through which this branch makes the structural transition from a single sheet at low $\Wi$ to the multilayered sheets seen at EIT.} 

\AS{To elaborate on this sheet shedding process, we focus on the attractor at $\Rey = 10000$, $\Wi = 4$, i.e., close to where this process first comes into existence}. \AS{Figures \ref{fig:SSP_Wi4}a to f cover one cycle of the process and correspond to $t = 4400$ to $4410$ taken every $2$ time units respectively.} \AS{The following description focuses on the bottom half of the channel with the same process taking place in the top half.}

\AS{As the name suggests, this process involves the shedding of sheets of polymer stretch from the primary TS structure, giving rise to multilayered sheets.} \AS{The onset of this process involves the arched portion of the TS structure seen between $x = 0$ and $x \approx 4$.} \AS{At the instant shown in (a), this arched section goes from a region of positive wall normal velocity to negative. The locations where the wall normal velocity changes sign is shown using green lines as seen near $x \approx 1.5$.}  \AS{Going from instant (a) to (c), this arch gets weaker and stretches out due to the wall normal velocity pulling the arch towards the centerline for $x \lesssim 1.5$ and towards the wall beyond this region.} \AS{Eventually, a piece of this arch breaks off, i.e., it is ``shed" as shown in (d) near the location where the wall normal velocity changes sign.} 

\AS{The next stage of the process has to do with these fragments that are shed, which get convected downstream faster than the primary TS structure. This can be seen in (e).} \AS{This difference has to do with the fact that the velocity of the primary TS structure is dictated by the mean velocity at the critical layer while these fragments are closer to the centerline, where the streamwise velocity is higher.} \AS{As these fragments are convected, they are simultaneously sheared, giving rise to the near-wall multilayered sheet seen in (f) between $x \approx 3$ and $x \approx 6$.} \AS{The instant (f) is similar to (a),} \MDGrevise{and the process repeats.}  \AS{The time scale of the process is dictated by the dominant time scale of the attractor with TS critical layer mechanisms at its core.} 

This process gets more pronounced as we move along this branch, and the statistical intermittency described earlier corresponds to a structural intermittency due to the sheet shedding process \AS{excited by viscoelasticity.}
\AS{To show this, Figure \ref{fig:SSP_Wi11} depicts the structural intermittency observed at $\Wi = 11$.} \AS{Snapshots shown are taken every 10 TU - the time scale at which we described the sheet shedding process at $\Wi = 4$.} The attractor goes from displaying a single sheet of polymer stretch as seen in (a) to strong multilayered sheets at the instant (j) and back. \AS{The observed steps of shedding-convecting-shearing now happen in a fashion that eventually gives rise to multilayered sheets.} These near-wall sheets observed during this intermittent process start to resemble those observed at EIT as described earlier. These multilayered sheets take over the dynamics at $\Wi = 12$, turning into EIT at $\Wi = 13$.


\begin{figure}
    \begin{center}
	\includegraphics[scale=0.4]{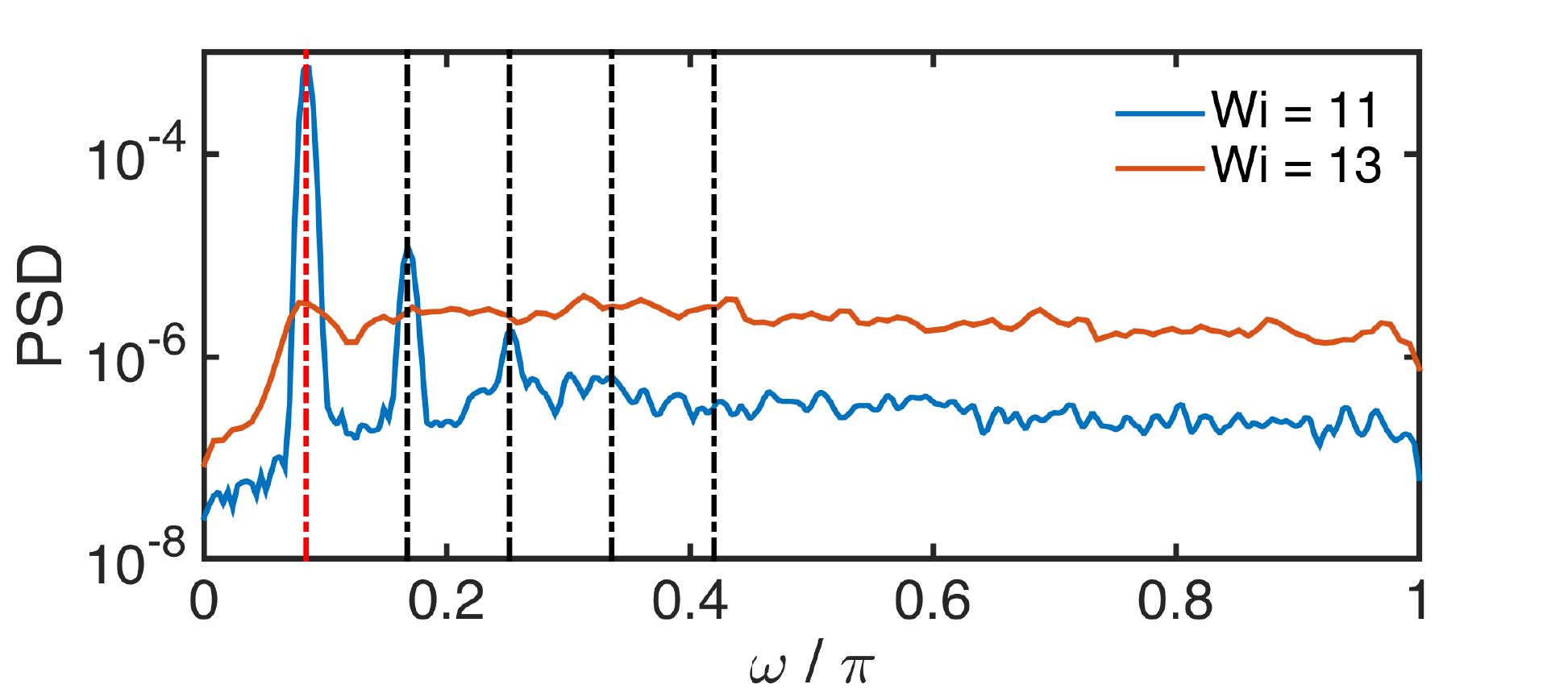}
			\caption[]{Power spectral density (PSD) of wall normal velocity at $y = \pm 0.8$ at the indicated $\Wi$.}
			\label{fig:Power_density}
    \end{center}
\end{figure}

To build on the evidence linking the TS attractor (up to $\Wi = $ 11) and EIT (beyond 11), Figure \ref{fig:Power_density} shows the power spectral density of wall normal velocity at $y = \pm 0.8$, with the following observations holding true at other $y$ locations. At $\Wi = 11$ (blue), the spectrum is mainly composed of the fundamental TS frequency (red-dashed) and its two higher harmonics (black-dashed). This is  unsurprising, as TS-like flow structures characterized by wall-normal velocity contour lines spanning the channel are a dominant feature of the attractor (evidenced by Figure \ref{fig:SSP_Wi11}). Interestingly, the fundamental TS frequency (red-dashed vertical line) remains a prominent feature of the spectrum even at $\Wi = 13$  (brown) in spite of the obvious structural differences with $\Wi = 11$. This points at TS-like mechanisms dictating the dynamics even at EIT.

\begin{figure}
	\begin{center}
		\begin{subfigure}{0.45\textwidth}
	     	\includegraphics[scale=0.4]{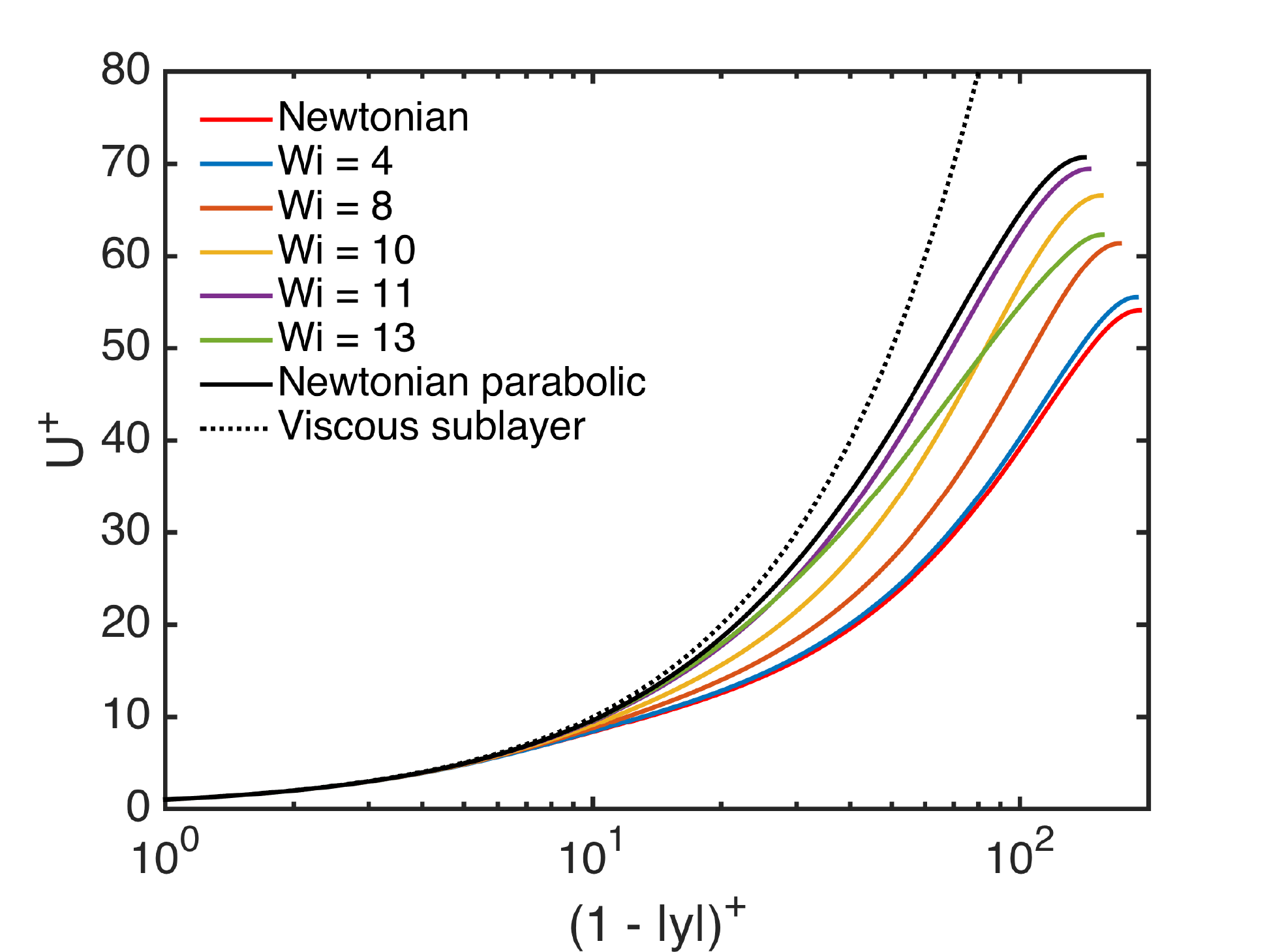} 
	     	\caption[]{}
      		\label{fig:Mean_vel}
        \end{subfigure}
		\begin{subfigure}{0.45\textwidth}
			\includegraphics[scale=0.4]{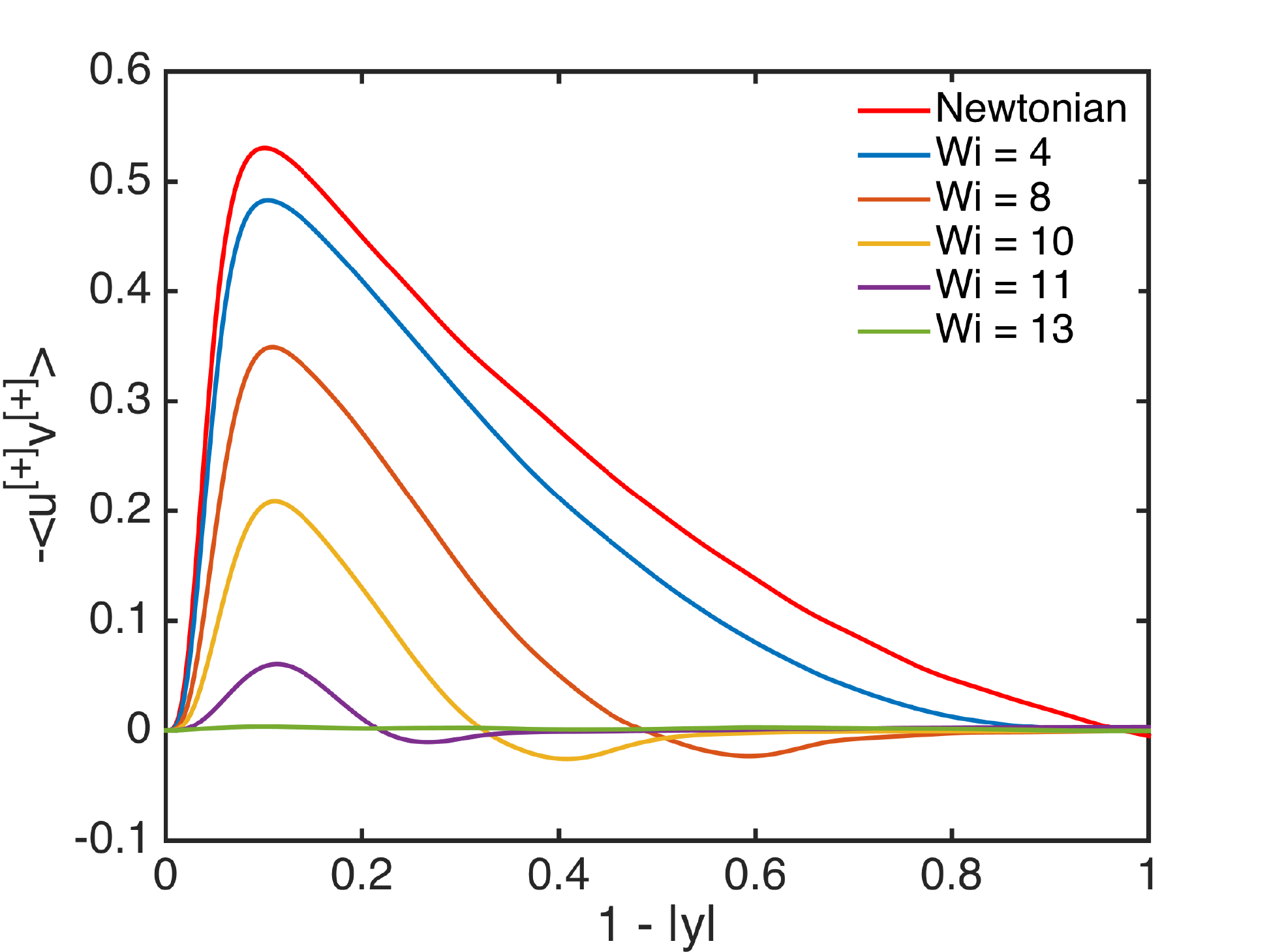} 
			\caption[]{}
			\label{fig:RSS}
		\end{subfigure}
		\begin{subfigure}{0.45\textwidth}
	    	\includegraphics[scale=0.4]{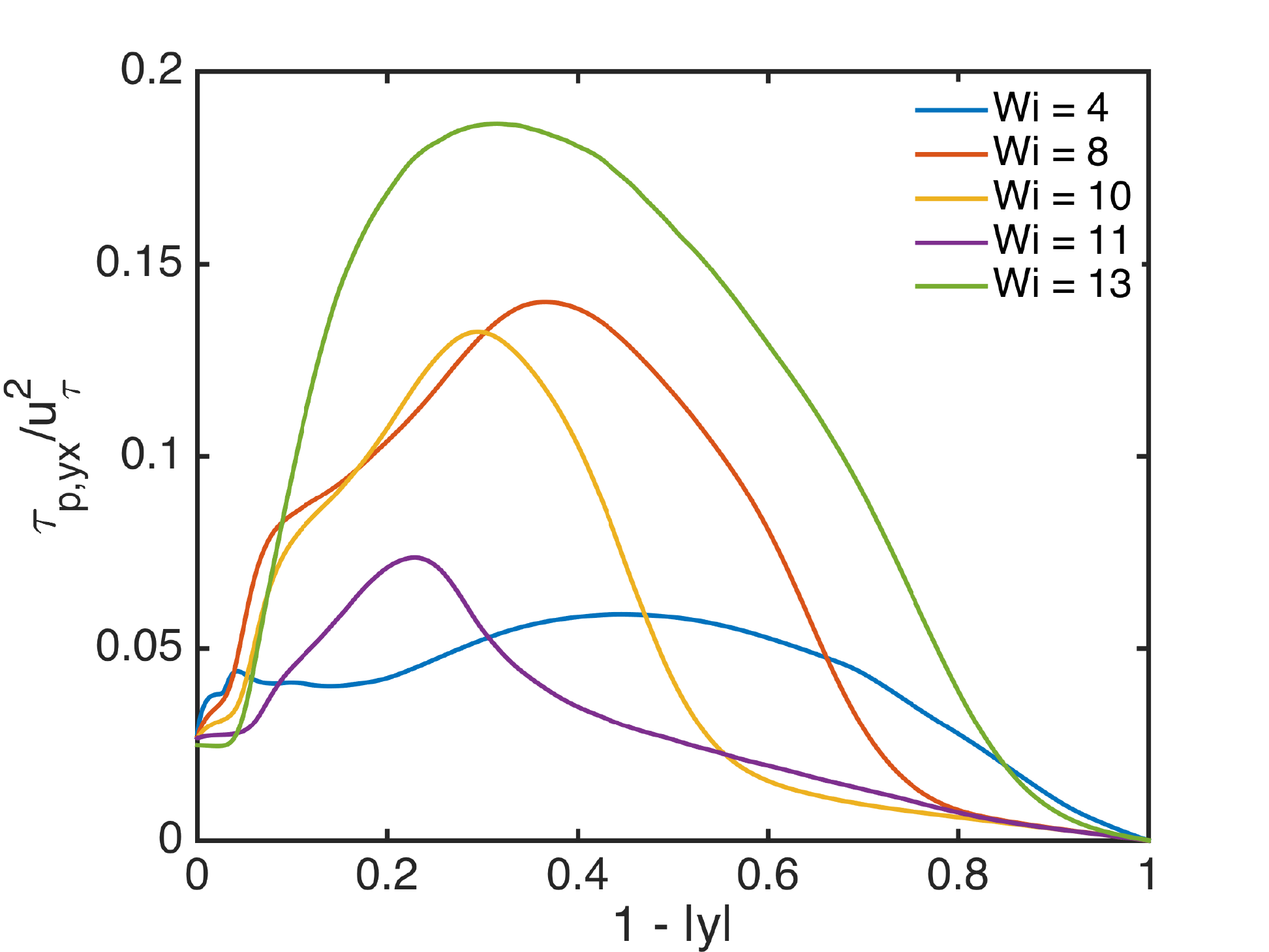}
	    	\caption[]{}
      		\label{fig:polymer_stress}
		\end{subfigure}
\caption[]{(a) Mean streamwise velocity profiles, (b) Reynolds shear stress profiles,  (c) polymer shear stress profiles of full space solutions. All profiles are shown as a function of distance from the wall $1 - |y|$. }
\label{fig:Mean_profiles}	     
\end{center}
\end{figure}

Looking at other statistics, Figure \ref{fig:Mean_profiles} shows the evolution of mean profiles of streamwise velocity (figure \ref{fig:Mean_profiles}a), Reynolds shear stress (figure \ref{fig:Mean_profiles}b) and polymer shear stress (figure \ref{fig:Mean_profiles}c) going from the Newtonian attractor to EIT. On increasing $\Wi$, the mean velocity profiles get increasingly uplifted from Newtonian (red) to $\Wi = 11$ (purple) with the profile at $\Wi = 4$ (blue) very similar to the Newtonian one. EIT at $Wi = 13$ (green) displays a profile in between Newtonian and $\Wi = 11$. Turning to the Reynolds shear stress profiles, the Newtonian profile (red) displays a near-wall peak corresponding to the critical layer of the TS wave. This points to a prominent role for critical layer behavior in Reynolds shear stress production. Moving along the branch, the attractors continue to display this near-wall peak with a monotonic decrease in magnitude until $\Wi = 13$, where the Reynolds shear stress becomes negligible. Past studies of EIT have reported reported similar observations \citep{Samanta:2013el, Sid:2018gh, Shekar:2019hq}. While Reynolds shear stress dominates over polymer shear stress at $\Wi = 4$, the roles are completely reversed at EIT, with polymer shear stress being the main source of turbulence generation. At $\Wi = 11$, the above quantities are of similar magnitude thus indicating roles of equal importance in self-sustenance. Further, the peak polymer shear stress at EIT loosely aligns with the peaks at lower $\Wi$. This hints at  mechanisms originating from the TS attractor underlying the main source of turbulence production at EIT. Taken together, these results provide direct indications of a connection between the TS attractor and EIT.

\subsection{Intermittent regime: analogy with relaxation oscillations\label{sec:shilnikov}}

\begin{figure}
	\begin{center}
		\begin{subfigure}{0.45\textwidth}
			\includegraphics[scale=0.4]{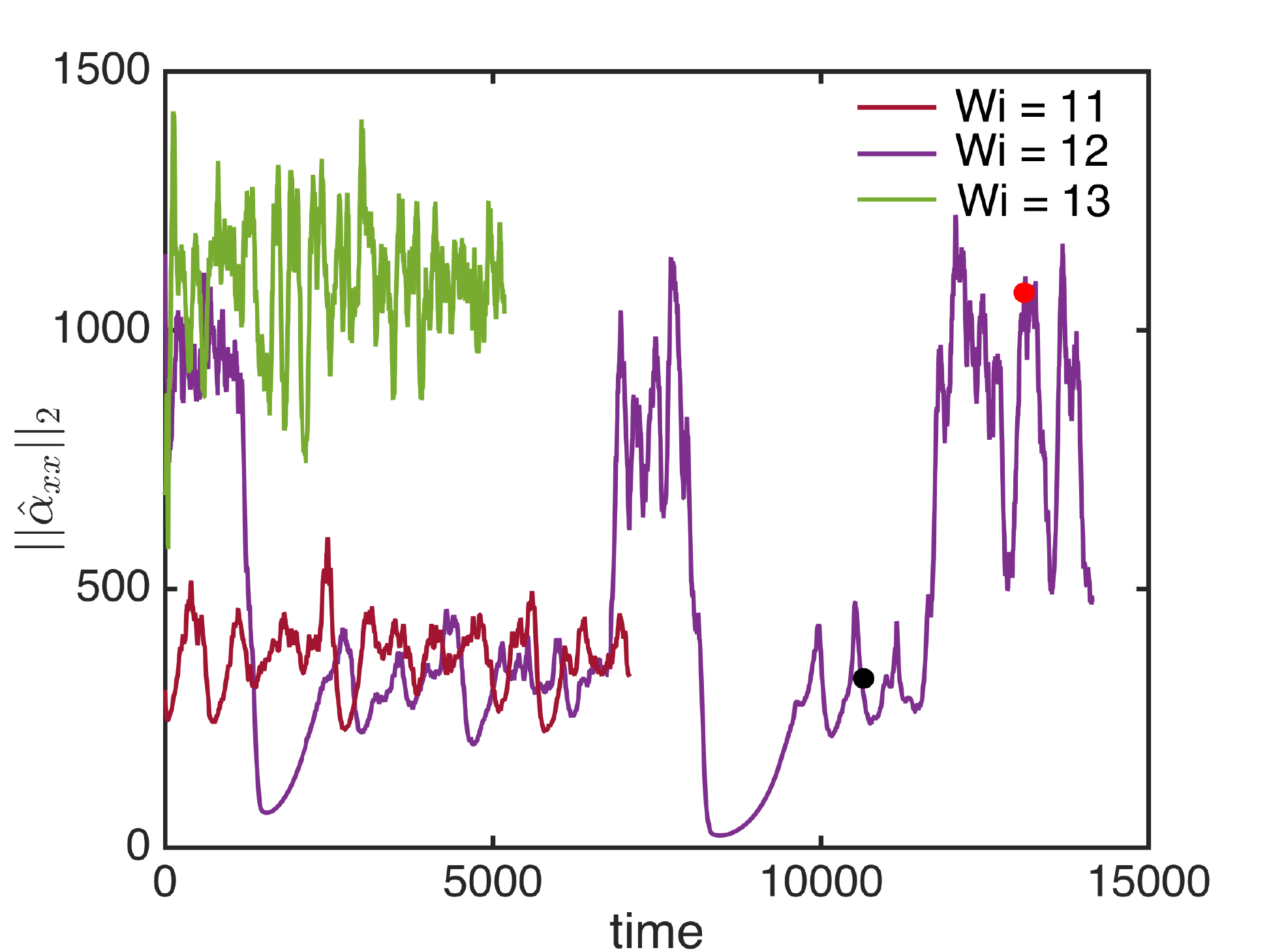}
			\caption[]{}
			\label{fig:L2_Cxxp_vs_time_Wi12_SR}
		\end{subfigure}\hspace{0.3\textwidth}
		\begin{subfigure}{0.45\textwidth}
	     	\includegraphics[scale=0.33]{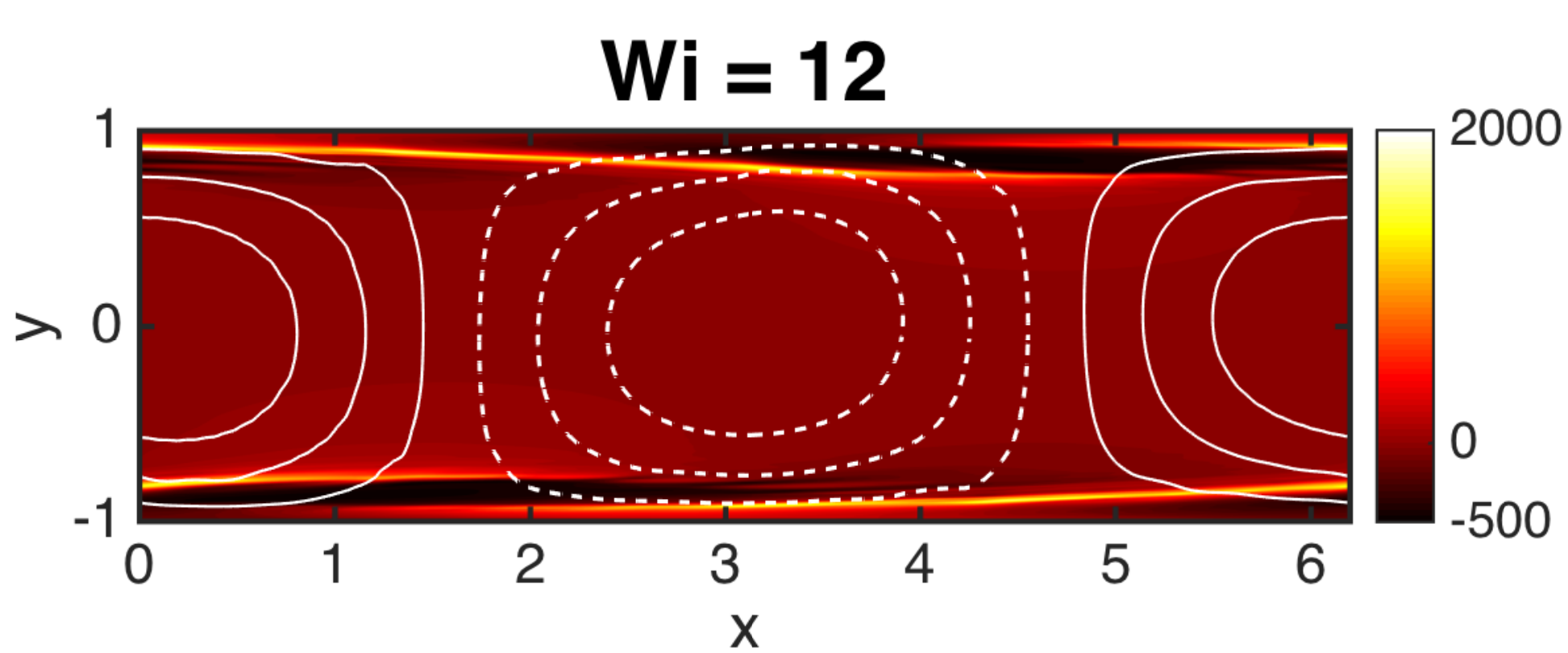} 
	     	\caption[]{}
      		\label{fig:Wi12_SR_TS}
        \end{subfigure}
		\begin{subfigure}{0.45\textwidth}
			\includegraphics[scale=0.33]{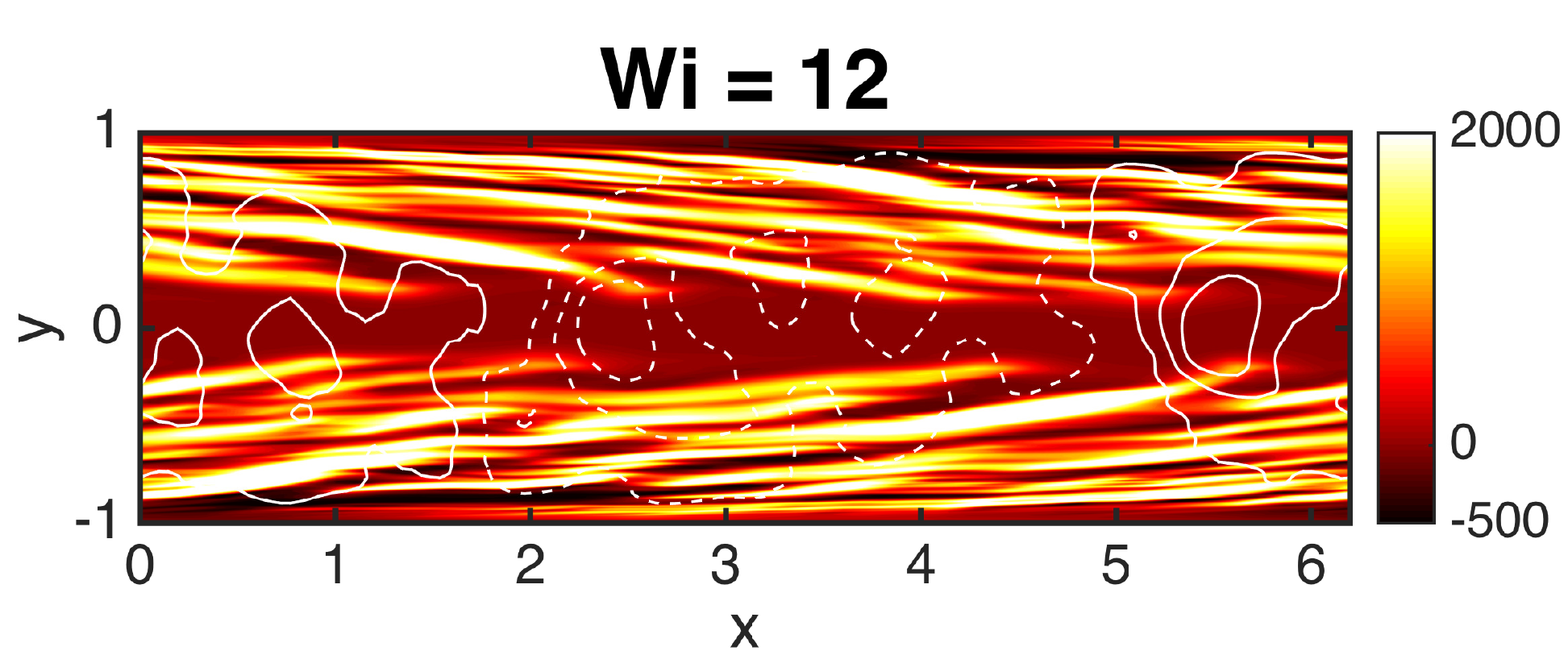} 
			\caption[]{}
			\label{Wi_12_SR_EIT}
		\end{subfigure}
\caption[]{(a) Instantaneous $||\hat{\alpha}_{xx}||_{2}$ vs time at $\Rey = 10000$ in the shift-reflect symmetric subspace. (b) and (c) are snapshots of the fluctuation structure of the TS (instant indicated by the black dot) and EIT (red dot) metastable states respectively.}
\label{fig:Re10000_SR}	     
\end{center}
\end{figure}

\MDGrevise{In the range $10\lesssim \Wi \lesssim 13$, the amplitude of $||\hat{\alpha}_{xx}||_{2}$ increases sharply and its temporal fluctuations are large. At the beginning of this regime, the flow remains structurally similar to the TS wave, but with substantial sheet-shedding (Figure \ref{fig:Re10000_bif_dia_FS}f, $\Wi=10$) while at the end is found the full-blown EIT structure, with sheets occupying most of the channel (Figure \ref{fig:Re10000_bif_dia_FS}i). To gain some understanding of this transition, we focus on the dynamics in the shift-reflect subspace.  With this restriction, the intermittency and basic structural features of the  full space simulations are found, while the symmetry-breaking that leads to the degenerate, strongly asymmetric structures like that in Figure \ref{fig:Re10000_bif_dia_FS}h will not complicate matters.   }


\MDGrevise{Figure \ref{fig:Re10000_SR}a shows the time evolution of the instantaneous value of $||\hat{\alpha}_{xx}||_{2}$ from simulations in the S-R subspace at $\Wi=11,12$, and $13$. While there are substantial fluctuations in the flows at $11$ and $13$, the behavior at $\Wi=12$ is quite distinct, displaying strong intermittency between two metastable states. A snapshot from the lower of these states (Figure \ref{fig:Re10000_SR}b) closely resembles a TS wave, while one from the upper (Figure \ref{fig:Re10000_SR}b) looks like EIT. There is no hysteresis in this scenario: if we use initial conditions at higher $\Wi$ to generate trajectories at lower $\Wi$, we get the same behavior as when we use initial conditions at lower $\Wi$ to generate trajectories at higher $\Wi$.  } 

\MDGrevise{The destabilization of one attractor (TS here) leading to a regime of intermittency followed by appearance of another attractor with substantially different amplitude is typical of a dynamical system with multiple time scales with a pleated slow manifold and the possibility of rapid jumps between different, locally attracting, pleats \citep{Guckenheimer:2004ub}.}
\MDGrevise{In particular, }\AS{ we draw an analogy with the following van der-Pol-type system:}
\begin{align}
        \label{vdp_1}
                \dot{x} = \frac{1}{\epsilon}(y - (x^3/3 - x)), \\
		\label{vdp_2}
                  \dot{y} = a - x                             
\end{align}

\begin{figure}
    \begin{center}
	\includegraphics[scale=0.4]{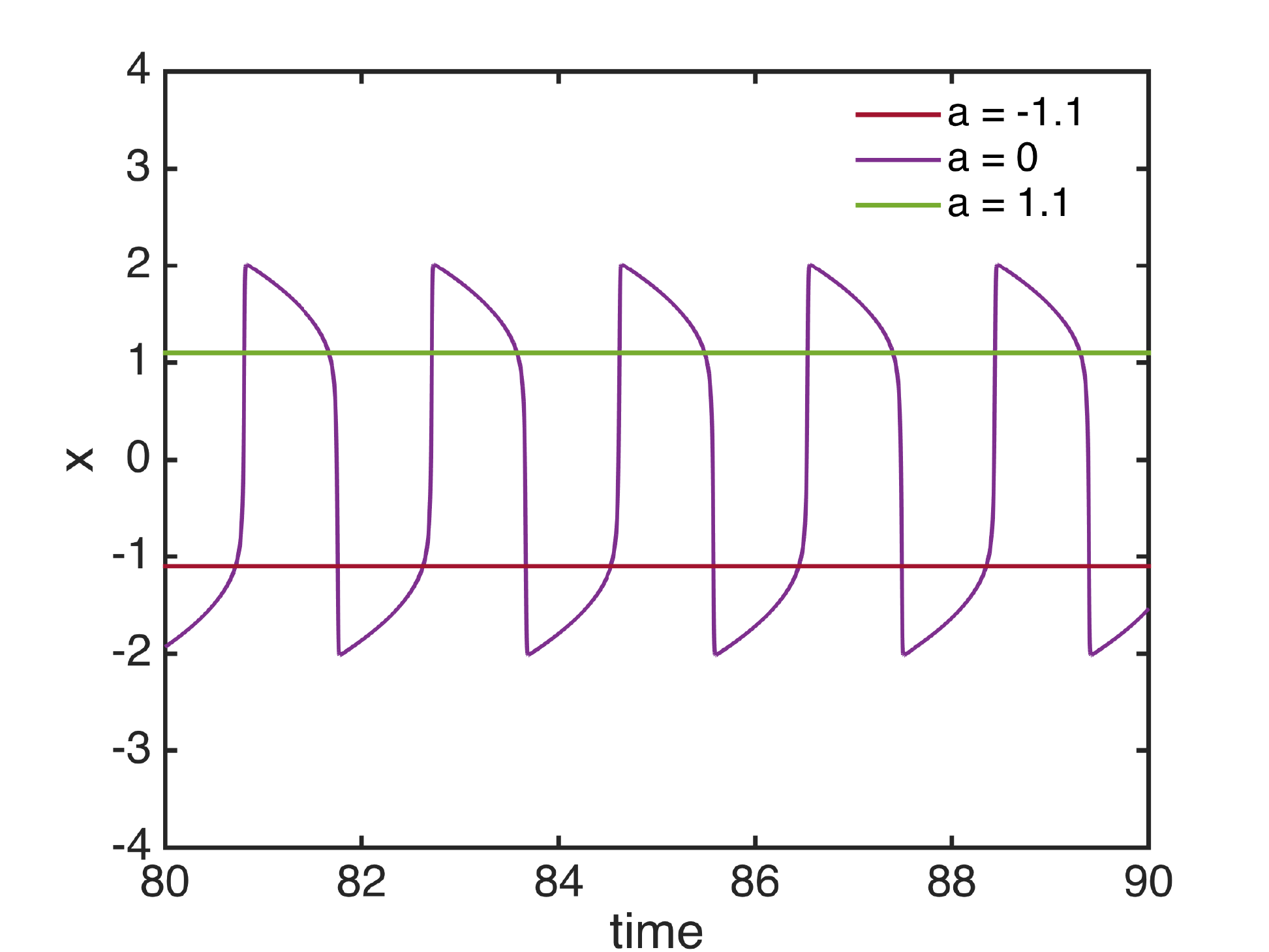}
			\caption[]{Dynamics of $x$ in the van der-Pol-type system described by Equations \ref{vdp_1} and \ref{vdp_2} for $\epsilon = 0.01$ at the indicated values of control parameter $a$.}
			\label{fig:Relaxation_Oscillation}
    \end{center}
\end{figure}

\AS{Here, $a$ is the control parameter of the system and $\epsilon$ determines the ratio of time scales between the $x$ and $y$ dynamics.} \MDGrevise{The classical van der Pol system is recovered when $a=0$. At each value of $a$, there is a unique steady state $(x,y)=(a,a^3/3-a)$. As $\epsilon\rightarrow 0$, this state is stable when $a<-1$ and $a>1$, while in $-1<a<1$, there is a stable limit cycle -- a relaxation oscillation that jumps rapidly between positive and negative values of $x$. The bifurcations are all supercritical so as in the above scenario there is no hysteresis.} \AS{Figure \ref{fig:Relaxation_Oscillation} shows the $x-$dynamics of the system for $\epsilon = 0.01$ at different values of $a$.} \AS{For $a = -1.1$ (maroon) and $a = 1.1$ (green), initial conditions settle on the stable steady state $x = -1.1$ and $x = 1.1$, respectively.} \AS{For $a = 0$ (purple), the system displays relaxation oscillations.} \AS{The observations presented in this section, taken in consideration with evidence of a link between TS and EIT, are entirely consistent with a high-dimensional version of such relaxation oscillations between TS and EIT metastable states at $\Wi = 12$, which give rise to stable EIT at higher $\Wi$.}

\subsection{Robustness in large spatial domains and very high polymer extensibility \label{sec:other_params}}

Finally, we present results in other parameter regimes to portray the robustness of our observations. For each of the below scenarios, we look at snapshots of the attractor for different levels of viscoelasticity.

\begin{figure}[t]
\begin{center}
\captionsetup{justification=raggedright}
		\begin{subfigure}{0.45\textwidth}
			\includegraphics[scale=0.36]{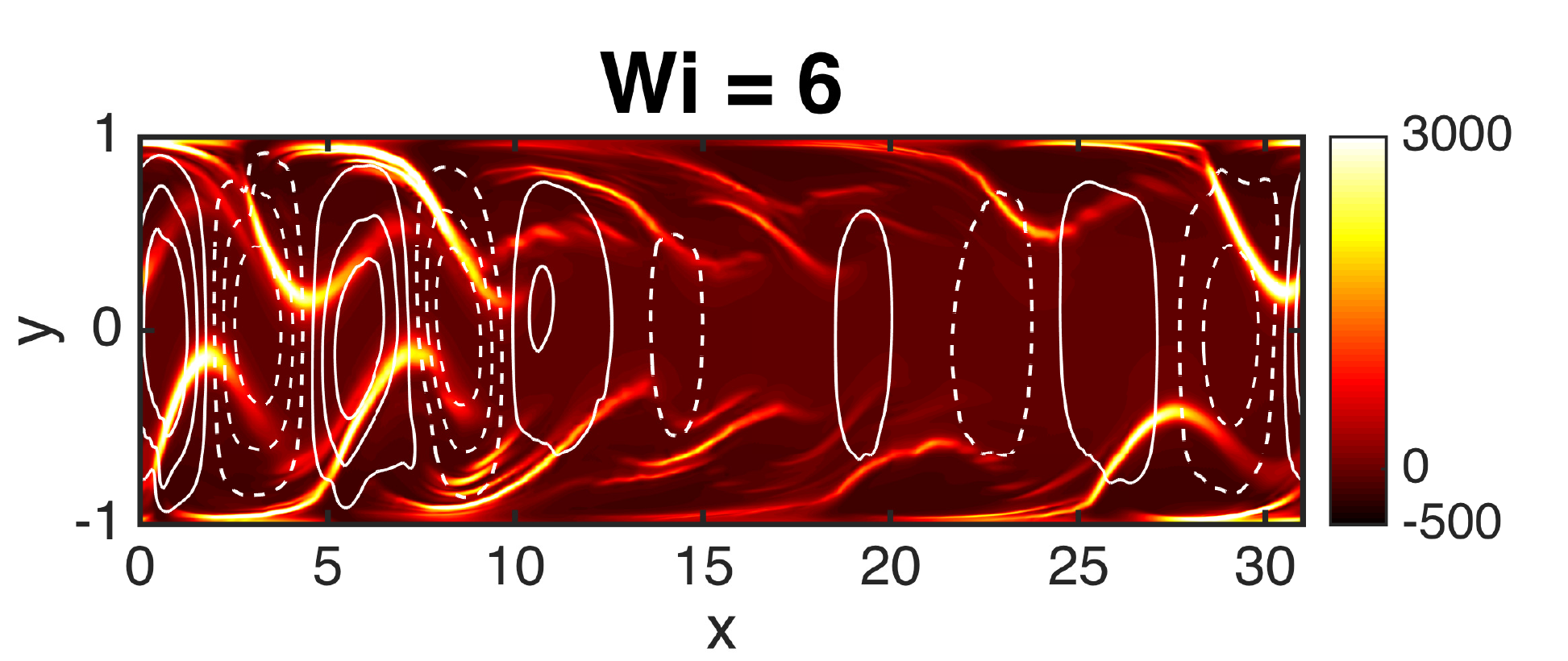}
			\caption[]{}
			\label{fig:Wi6_31}
		\end{subfigure}
		\begin{subfigure}{0.45\textwidth}
	     	\includegraphics[scale=0.36]{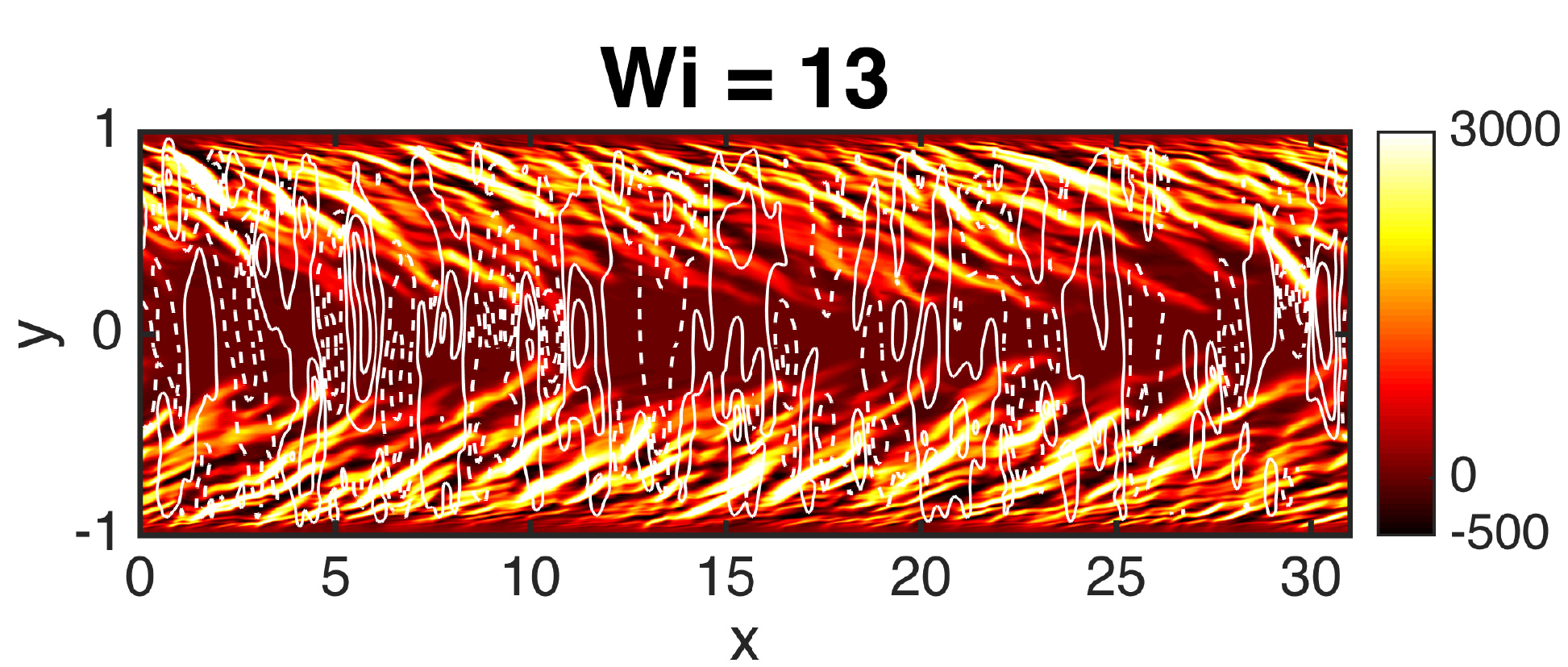} 
	     	\caption[]{}
      		\label{fig:Wi13_Lx31}
        \end{subfigure}


\caption[Snapshots of TS and EIT attractors in large domains]{Snapshots of the attractor at $\Rey = 10000$ in a box of size $L_x = 31$ at the indicated $\Wi$. Contour plots follow the same format as previous figures. Note the compressed scale in $x$.}
\label{fig:Large_box}
\end{center}	     
\end{figure}

First, we consider results at $\Rey = 10000$ in a large domain of size $L_x = 31$, i.e., five times the domain size used above. Snapshots of the attractor are shown in Figure \ref{fig:Large_box}. At $\Wi = 6$, the attractor displays a structure corresponding to five spatial periods of the TS attractor in $L_x = 6.2$, with near-wall sheets and wall normal velocity across the channel. Further, we can see clear evidence of the sheet shedding process at play. Here, different parts of the domain can be at different stages of the process: i.e., there is  spatio-temporal intermittency. For example, we see multilayered sheets near $x = 10$ and sheet shedding near $x = 25$. At $\Wi = 13$, we see EIT with strong multilayered sheets of polymer stretch as we did for $L_x = 6.2$.

\begin{figure}
\begin{center}
		\begin{subfigure}{0.45\textwidth}		
			\includegraphics[scale=0.36]{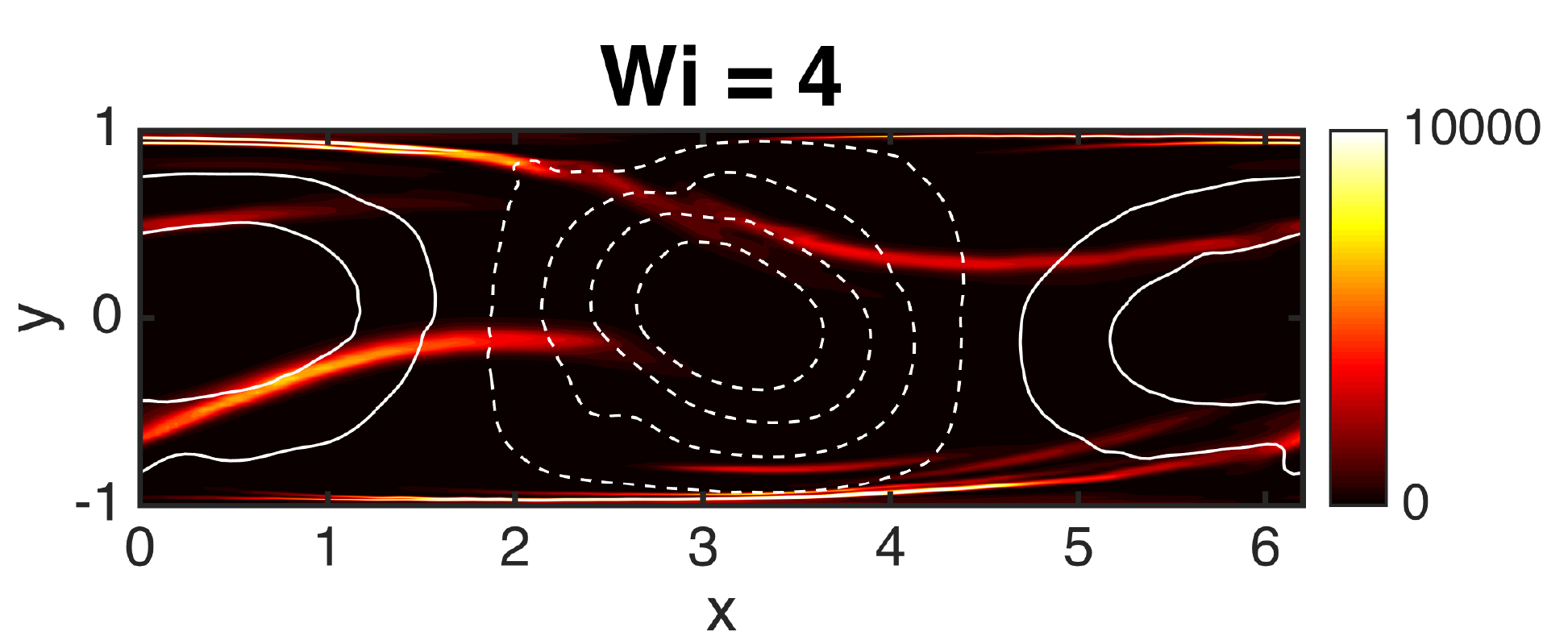}
			\caption[]{}
			\label{fig:Wi4_b100000}
		\end{subfigure}\hspace{0.3\textwidth}		
		\begin{subfigure}{0.45\textwidth}
	     	\includegraphics[scale=0.36]{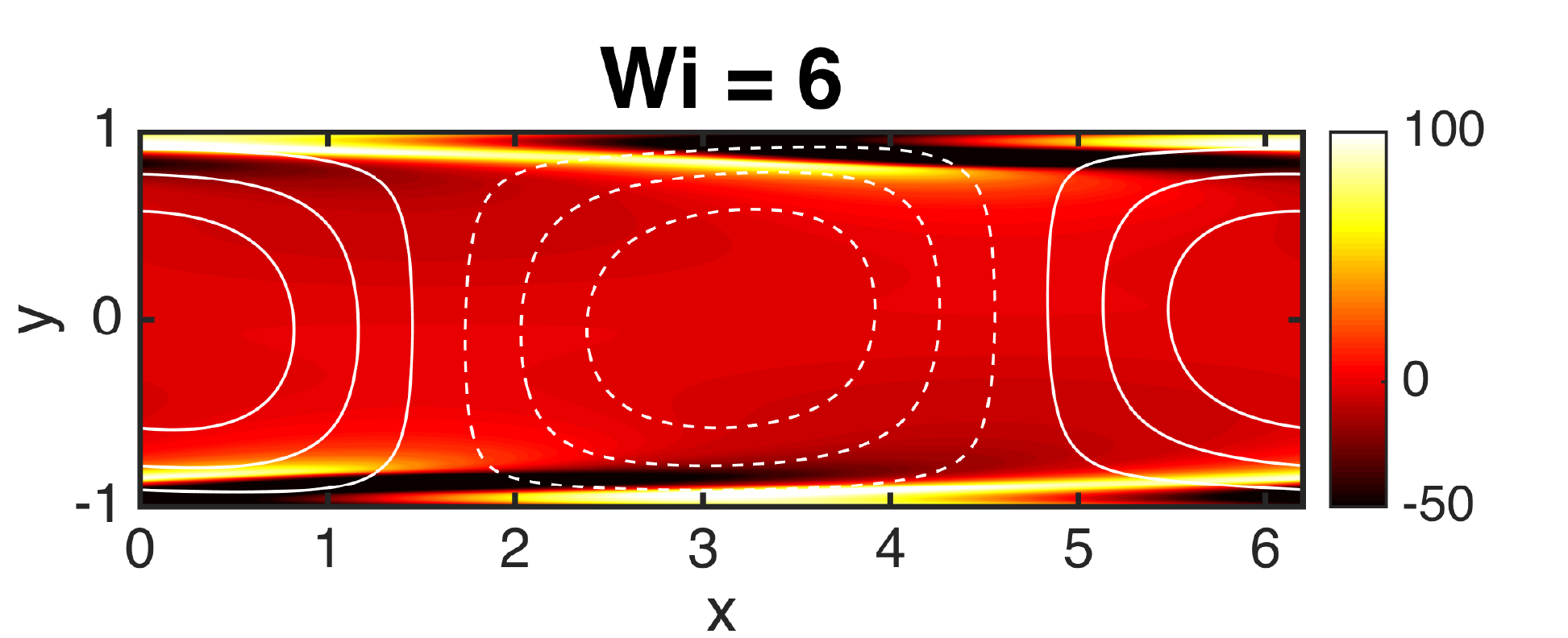} 
	     	\caption[]{}
      		\label{fig:Wi6_b100000_1}
      	\end{subfigure}
		\begin{subfigure}{0.45\textwidth}
	     	\includegraphics[scale=0.36]{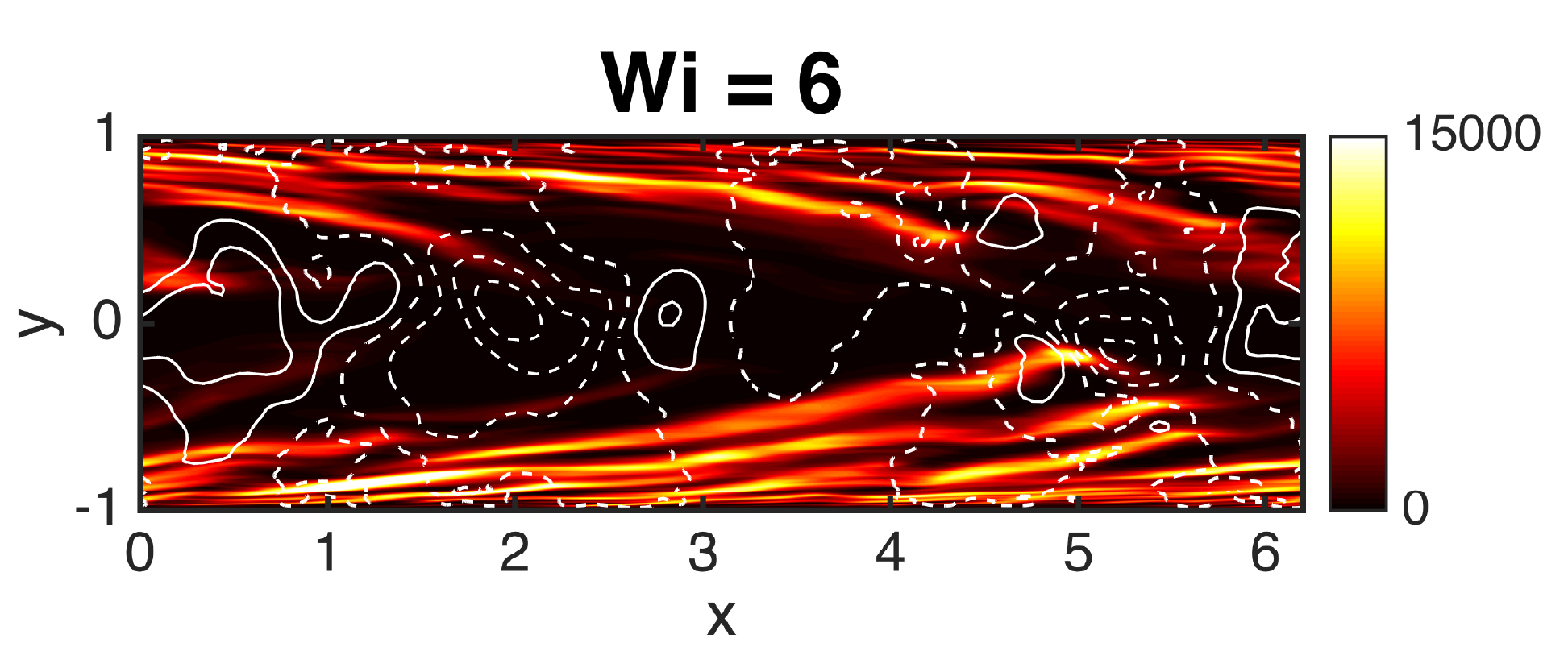} 
	     	\caption[]{}
      		\label{fig:Wi6_b100000_2}
        \end{subfigure}\hspace{0.3\textwidth}       
		\begin{subfigure}{0.45\textwidth}
			\includegraphics[scale=0.36]{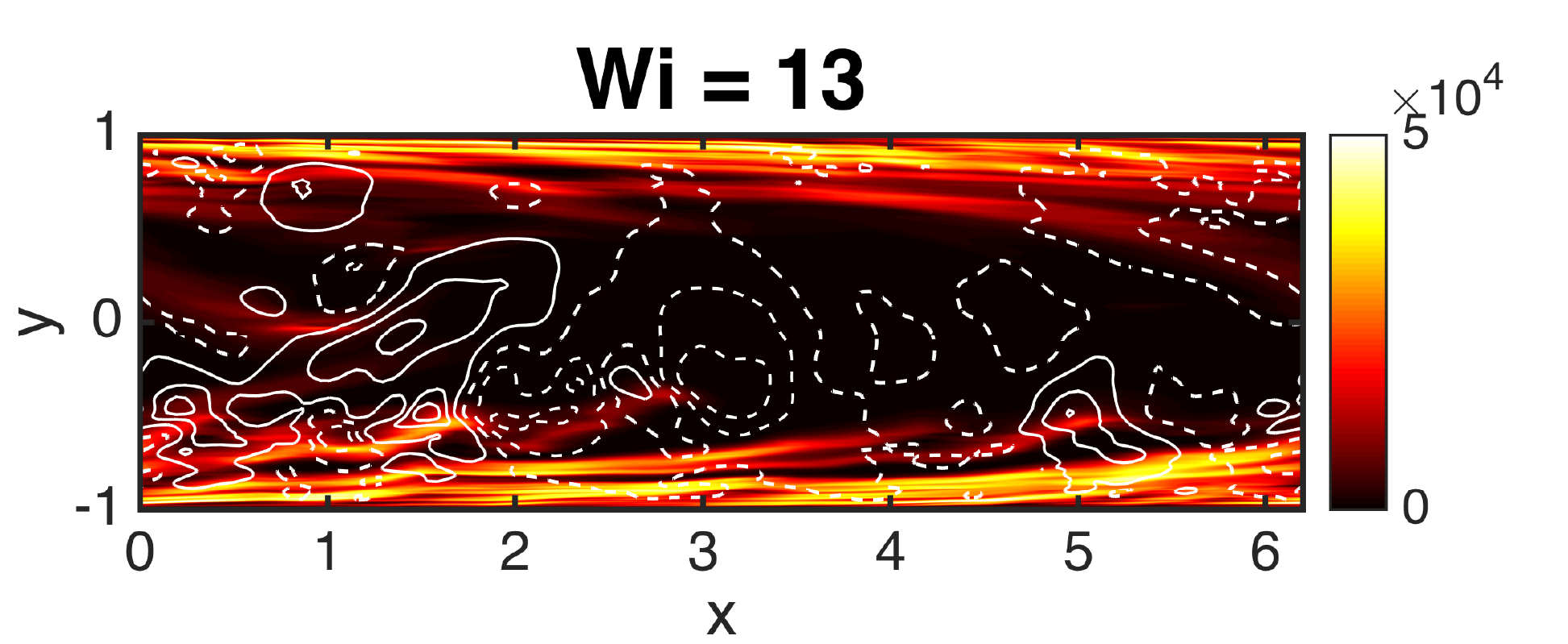} 
			\caption[]{}
			\label{Wi13_b100000}
		\end{subfigure}
		
\caption[Snapshots of the attractor at very high extensibility]{Snapshots of the attractor at $\Rey = 10000$ with $b = 100,000$ at various $\Wi$. At $\Wi=6$, the dynamics are strongly intermittent; (b) and (c), respectively, are snapshots from the low-amplitude (TS) and large amplitude (EIT) intervals. Format is the same as previous plots.}
\label{fig:high_b}
\end{center}	     
\end{figure}

Now we consider the scenario at very high extensibility, $b =  100,000$, with other parameters remaining the same. 
 Figure \ref{fig:high_b} depicts snapshots of the attractor for this case. 
At $\Wi = 4$, the attractor corresponds to the TS attractor in this parameter regime and displays the same characteristics as the attractor at lower extensibility. Further, the attractor continues to exhibit the sheet shedding process that leads to multilayered sheets of polymer stretch. This can be seen near $x = 4$. On increasing $\Wi$, the TS attractor loses stability before giving rise to stable EIT. At $\Wi = 6$, the dynamics alternate between TS and EIT metastable states as shown in Figures \ref{fig:high_b}b and c, respectively. \AS{This EIT metastable state can sometimes be asymmetric and thus in agreement with the observations at $\Wi = 12$, $b = 6400$.} These results are consistent with the relaxation oscillation scenario above \AS{with the bifurcation happening around $\Wi \approx 6$ at this value of extensibility}. At $\Wi = 13$,  we observe EIT.

\subsection{Discussion: flow geometry and dimensionality\label{sec:broader-context}}

\MDGrevise{As with our prior study, Ref.~\citep{Shekar:2020gt}, the present work is limited to two-dimensional flow in a plane channel. Given these limitations, it is important to consider how they may be viewed in the broader context of turbulent drag reduction, first with regard to how they may relate to the pipe flow geometry and then in the full context of three-dimensional turbulence. These issues were discussed at some length in Ref.~\cite{Shekar:2020gt}, so here we summarize that discussion and describe some further issues in connection to very recent literature. }

As noted in the Introduction, elastoinertial turbulence with very similar features has been observed in channel, pipe and plane Couette flows. A natural question, then, is to what extent the above channel flow results are relevant to the pipe flow case. 
While it is true of course that there is no linear instability of Newtonian pipe or plane Couette flows, there are nevertheless ``wall modes" with critical layer structure related to that of the TS mode \citep{DrazinReid}.
 Additionally, critical layers are a strong source of linear amplification in those flows, as they are in channel flow \citep{Mckeon:2010ep}. 
 Furthermore, the absence of a \emph{Newtonian} mechanism for nonlinear sustainment of Tollmien-Schlichting-like traveling waves does not imply the absence of a \emph{viscoelastic} mechanism. Indeed, Shekar et al.~\cite{Shekar:2020gt} showed the existence of just such a mechanism. Furthermore, as noted in the Introduction, pipe flow  \citep{lopez2019dynamics} and plane Couette flow \citep{pereira2019beyond} simulations at EIT display  essentially two-dimensional velocity fluctuations localized near the wall that are similar to those reported in channel flow, and resolvent analysis of viscoelastic pipe flow  demonstrates strong linear amplification of modes with near-wall critical-layer stress fluctuations \cite{Zhang:2021ef}.

 At the same time, in more strongly viscoelastic regimes, Garg et al.~\cite{garg2018} and Khalid et al.~\cite{Khalid:2021ix} have found a center mode instability for pipe flow and channel flow respectively, in the Oldroyd-B limit. To put these results into context,  for $\Rey=10000$, this instability comes into existence for $\Wi \gtrsim 500$ for $\beta = 0.97$ and $\Wi \gtrsim 200$ for $\beta = 0.9$ in either geometry. \cite{page2020exact} numerically converged nonlinear ``arrowhead'' structures originating subcritically from this instability in channels at very low $\Rey \sim O(100)$, $\Wi \sim O(10)$. Similar structures were observed in 2D simulations of EIT by Dubief et al.~\cite{dubief2020first} at very high viscoelasticity with $\Wi \sim O(100)$. Choueiri et al.~\cite{choueiri2021experimental} also note the appearance of ``chevron" shaped structures resembling the unstable center mode in pipe flow at very low $\Rey \sim O(100)$ before being taken over by near-wall modes on increasing $\Rey$. These results open up the possibility that other states unrelated to the nonlinear excitation of a wall mode may also play a role at EIT in both channels and pipes, especially at very low $\Rey$ and high $\Wi$. Nevertheless, the present work demonstrates a direct connection between a Newtonian wall mode (the TS mode) and two-dimensional EIT structures.

We now turn to the topic of how the present results are related to the fully three-dimensional context of near-wall turbulence. Newtonian turbulence is of course, strongly three-dimensional, with the dominant near-wall structure comprised of coherent wavy streamwise vortices that to some extent can be modeled by so-called Exact Coherent States (ECS) \citep{Graham:2020ba}.
In particular, Li et al.~\cite{Li:2007ii} studied the bifurcation scenario for a particular family of channel flow ECS (\cite{Waleffe:1998wk}) in a parameter regime close to that considered here. At $\Rey=1500$, this ECS family is sufficiently weakened by viscoelasticity to lose existence at $\Wi\approx 16$, somewhat above the value $\Wi\approx 10$ beyond which Newtonian turbulence cannot self-sustain in the DNS study of \cite{Shekar:2019hq}. (This discrepancy is consistent with what we know about transition in the Newtonian case: channel flow turbulence is self-sustaining above $\Rey\approx 1000$, while ECS can exist in that case down to $\Rey\approx 660$ \citep{Shekar:2018gg}.) Extrapolating slightly from the results of Li et al.~\cite{Li:2007ii}, one can estimate that at $\Rey=10000$, this ECS family exists at least up to $\Wi = 25$. Based the results above, we might expect that  $\Wi\gtrsim 13$, a mix of EIT-like and quasistreamwise structures might coexist, while once $Wi>25$, two-dimensional EIT would dominate.  
Consistent with this analysis, our preliminary work \citep{Shekar:2021vl} as well as a number of other recent studies have reported coexistence of EIT-like sheets  and 3D quasistreamwise structures \citep{Dubief:2013hh, Choueiri:2018it,  pereira2019beyond, pereira2019common, zhu2021nonasymptotic, wenhua2021role}. 
How the 2D and 3D structures interact is an important topic for future work.

\section{Conclusion}


This study describes the evolution of the Tollmien-Schlichting (TS) attractor in $\Rey- \Wi$ parameter space and its relationship to elastoinertial turbulence (EIT), using simulations of 2D channel flow of a very dilute polymer solution. \MDGrevise{At $\Rey=3000$, there is a solution branch with TS-wave structure but which is not connected to the Newtonian solution branch.}  At fixed Weissenberg number, $\Wi$ and increasing Reynolds number, $\Rey$ from 3000-10000, \MDGrevise{this} attractor goes from displaying a
striation of \MDGrevise{weak} polymer stretch localized at the critical
layer as described in Shekar et al.~\cite{Shekar:2020gt} to an extended sheet of \MDGrevise{very large} polymer stretch. We show that this transition is directly tied to the strength of the TS critical layer fluctuations and can be attributed to a coil-stretch transition when the local $\Wi$ at the hyperbolic stagnation point \MDGrevise{of the Kelvin cat's eye structure of the TS wave} exceeds $\frac{1}{2}$. \MDGrevise{At $\Rey=10000$, unlike $3000$, the Newtonian TS attractor evolves continuously into  EIT as $\Wi$ is increased from zero to about $13$ -- the two flows are directly connected in parameter space. We describe how the structure of the flow and stress fields changes with $\Wi$, highlighting in particular a ``sheet-shedding" process by which the individual sheets associated with the critical layer structure break up to form the layered multisheet structure characteristic of EIT.  Furthermore, in an intermediate parameter regime a state of highly intermittent dynamics are observed: the flow behavior alternates between a state with sheetlike stress fluctuations localized near the wall -- the TS structure --  to one with such structures present across the channel, except near the centerline -- EIT.  To understand this regime, we draw analogies with relaxation oscillator dynamics.  } 

The linear instability to Tollmien-Schlichting waves does not arise for pipe or plane Couette flow, so the scenario described here does not directly apply to those geometries. On the other hand, in these geometries elastoinertial turbulence with very similar features does arise in simulations in the same general parameter regime: namely, fluctuations localized in a layer near the wall, with a sheetlike stress structure and little to no activity in the center of the flow \cite{lopez2019dynamics,pereira2019beyond}. Furthermore, while linearly stable, wall modes analogous to the TS wave do exist in these other geometries, and may be subject to nonlinear critical layer excitation, and subsequent evolution into EIT, just as the TS mode is in the channel flow case.

\section*{Acknowledgments}
This work was supported by (UW) NSF  CBET-1510291, AFOSR  FA9550-18-1-0174, and ONR N00014-18-1-2865, and (Caltech) ONR N00014-17-1-3022.

Declaration of interests: none.


%

\end{document}